\documentclass[
 amsmath,amssymb,
 aps,twocolumn,]{revtex4-2}
\usepackage{graphicx} 
\usepackage{dcolumn}
\usepackage{bm}
\usepackage{amsmath}
\usepackage{epsfig}
\usepackage{color}

\begin{document}
\title{First-passage time of a Brownian searcher with stochastic resetting to random positions}

\author{V. M\'endez, R. Flaquer-Galm\'es, D. Campos  }
\affiliation{Grup de F\'{\i}sica Estad\'{\i}stica, Departament de F\'{\i}sica. Facultat de Ci\`{e}ncies, Universitat Aut\`{o}noma de Barcelona, 08193 Barcelona, Spain.}

\begin{abstract}
We study the effect of a resetting point randomly distributed around the origin on the mean first passage time of a Brownian searcher moving in one dimension. We compare the search efficiency with that corresponding to reset to the origin and find that the mean first passage time of the latter can be larger or smaller than the distributed case, depending on whether the resetting points are symmetrically or asymmetrically distributed. In particular, we prove the existence of an optimal reset rate that minimizes the mean first-passage time for distributed resetting to a finite interval if the target is located outside this interval. When the target position belongs to the resetting interval or it is infinite then no optimal reset rate exists, but there is an optimal resetting interval width or resetting characteristic scale which minimizes the mean first-passage time. We also show that the first-passage density averaged over the resetting points depends on its first moment only. As a consequence, there is an equivalent point such that the first-passage problem with resetting to that point is statistically equivalent to the case of distributed resetting. We end our study by analyzing the fluctuations of the first-passage times for these cases. All our analytical results are verified through numerical simulations.      
\end{abstract}

\maketitle

\section{Introduction}
The theory of resetting processes has been developed considerably in the last decade \cite{EMS20,EM2011,KMSS14,P15,CM15,BBR16,PKE16,Re16,EM18,NG23} and has allowed to gain understanding in a variety of topics such as the optimization of enzymatic reaction kinetics \cite{RUK14,RRU15,RUK14b}, computational searches \cite{MZ02} or animal foraging \cite{BS14,PKR20,V22}. The paradigmatic case of Brownian motion under stochastic resetting exhibits first passage features different from ordinary diffusion \cite{EM2011prl,EM2011,Re16,PaRe17}. When the Brownian searcher is randomly reset to a fixed position with a constant reset rate, the mean first passage time (MFPT) to a static target becomes finite and can be minimized with respect to the reset rate \cite{EM2011prl}. Resetting, however, is not necessarily beneficial in all cases, so a general criterion can be established to determine when it can effectively reduce the MFPT of a random walker to a static target \cite{Re16,PR17,CS18}. Furthermore, the fluctuations of the optimal first passage time (when it exists) are found to share remarkably simple universal properties \cite{Re16,Be20}.

Most of the current knowledge in the field strongly relies on the case where a random searcher resets its position to a single point. Despite their relevance, the phenomenology of processes where the resetting point can vary from one reset to another is far less known. Distributed resetting was discussed formally for Brownian motion in the seminal paper \cite{EM2011}. Since then, some works have gone further and have considered the case where the resetting position is random and drawn from a distribution.  It is also highly natural to assume that in experiments or nature resets cannot be performed with perfect precision. In addition, from a theoretical perspective, the resetting distribution introduces new length scales into the problem, which can give rise to interesting phenomena \cite{SDB23}. Recently, it has been considered Poissonian resetting to multiples nodes in a network \cite{Go21}, or the thermodynamical aspects of distributed resetting \cite{TB23,MOSK23}. However, studies on the fundamental properties of distributed resetting points are still scarce \cite{O23}. 

Here we analyze the MFPT of a Brownian searcher moving in one dimension when it resets its position at a constant rate to randomly distributed points $x_0$ according to a probability density function (PDF) $f(x_0)$. This PDF may account for a known territory that is revisited as a consequence of spatial memory. Resetting to this territory is a consequence of site fidelity that many terrestrial and marine species display to foraging and/or breeding locations \cite{Ca20,Mo09}. We also compare the efficiency (MFPT) and performance (coefficient of variation) of the search process between resetting to the origin or resetting to randomly distributed points.

The paper is organized as follows. In section II we obtain an exact analytic expression for the two first moments of the first passage time PDF for the aforementioned situation.  In section III we focus on the case where the resetting points are symmetrically distributed in an interval and the target is located outside. In section IV we consider the case where the target is located inside the resetting interval, and the case when the resets are distributed along the whole real line. In section V we consider the case of asymmetric resetting points. Next, in section VI we study the properties of the Equivalent Resetting Point, for which resetting to that point results in the same MFPT than for the case of distributed resets to an interval. In section VII we study the corresponding fluctuations of the first passage time for the most interesting cases above. Finally, in section VIII we conclude with a brief recapitulation of our results.

\section{MFPT}
Consider a random walker such that at constant rate $r$ its location resets to a random position $x_0$ drawn from the PDF $f(x_0)$. The trajectory of the searcher is terminated whenever it hits the target position, $x_T$. The first-passage problem for this situation states that the Master equation for the probability of finding the searcher at point $x$ at time $t$ is given by

\begin{equation}
    \frac{\partial P}{\partial t}=D\frac{\partial^{2}P}{\partial x^{2}}-rP+rf(x)Q(t)-p_{f}(t)\delta(x-x_{T}),
    \label{me}
\end{equation}
where $D$ is the diffusion coefficient, $Q(t)$ is the survival probability up to time $t$ and the rightmost term represents the (absorbing) boundary condition $P(x=x_T,t)=0$, with $p_f(t)$ the first passage time PDF. The latter is related to the survival probability via
\begin{equation}
    Q(t)=1-\int_0^t p_f(t')dt'.
    \label{Q}
\end{equation}
Note that in (\ref{me}) we multiply by $Q(t)$ the term corresponding to the searchers diffusing after resetting, to ensure explicitly that it includes only those trajectories that are not terminated (or absorbed) by the effect of the sink term $p_{f}(t)\delta(x-x_{T})$.

To solve \eqref{me} we perform the Fourier-Laplace (FL) transform with the initial condition $P(x,0)=\delta (x-x_0)$, where $x_0$ is a random variable. The FL transform of $P(x,t)$ is defined by
$$
P(k,s)=\int_{0}^{\infty}dse^{-st}\int_{-\infty}^{\infty}dxe^{ikx}P(x,t),
$$
so that FL transforming \eqref{me} we find
\begin{eqnarray}
    P(k,s)=\frac{e^{ikx_{0}}-e^{ikx_{T}}}{s+r+Dk^{2}}+\frac{rf(k)+se^{ikx_{T}}}{s+r+Dk^{2}}Q(s),
    \label{pks}
\end{eqnarray}
where we made use of the Laplace transform of \eqref{Q}, i.e, $p_f(s)=1-sQ(s)$, and $f(k)$ and $Q(s)$ are the Fourier and Laplace transforms of the resetting positions PDF and the survival probability, respectively. 
Now we need to impose the absorbing boundary condition at the target position. Working on the real space for the space positions and the Laplace space for time, and using the definition of the inverse Fourier transform we may write
$$
P(x,s)=\frac{1}{2\pi}\int_{-\infty}^{\infty}e^{-ikx}P(k,s)dk,
$$
hence, the boundary condition $P(x=x_T,s)=0$
turns into
$$
\int_{-\infty}^{\infty}e^{-ikx_{T}}P(k,s)dk=0.
$$
In consequence, multiplying \eqref{pks} by $e^{-ikx_{T}}$, integrating over $k$ and equating to zero we find
\begin{eqnarray}
   Q(s)=\frac{\int_{-\infty}^{\infty}\frac{1-e^{ik(x_{0}-x_{T})}}{s+r+Dk^{2}}dk}{\int_{-\infty}^{\infty}\frac{rf(k)e^{-ikx_{T}}+s}{s+r+Dk^{2}}dk}.
    \label{eqQ}
\end{eqnarray}
The integrals involved in \eqref{eqQ} can be computed as follows:
$$
\int_{-\infty}^{\infty}\frac{dk}{s+r+Dk^{2}}=\frac{\pi}{\sqrt{D(s+r)}}$$
$$\int_{-\infty}^{\infty}\frac{e^{ik(x_{0}-x_{T})}}{s+r+Dk^{2}}dk=\frac{\pi e^{-|x_{0}-x_{T}|\sqrt{\frac{r+s}{D}}}}{\sqrt{D(s+r)}},
$$
so finally through the convolution property of the Fourier transform it follows that
\begin{eqnarray}
  & &\int_{-\infty}^{\infty}\frac{f(k)e^{-ikx_{T}}}{s+r+Dk^{2}}dk\nonumber\\
  &=&\frac{\pi}{\sqrt{D(s+r)}}\int_{-\infty}^{\infty}f(x_{0})e^{-|x_{T}-x_{0}|\sqrt{\frac{r+s}{D}}}dx_{0}.  
\end{eqnarray}
Inserting these results into \eqref{eqQ}, we get the explicit expression for the survival probability in the Laplace space:
\begin{equation}
   Q(s)=\frac{1-e^{-|x_{0}-x_{T}|\sqrt{\frac{r+s}{D}}}}{s+r\int_{-\infty}^{\infty}f(x_{0})e^{-|x_{T}-x_{0}|\sqrt{\frac{r+s}{D}}}dx_{0}}.
    \label{Q2}
\end{equation}
Note that the small $s$ expansion for the PDF of the MFPT is
\begin{eqnarray}
  p_{f}(s)=1-sT+\frac{s^{2}}{2}T^{2}+...
  \label{exppf}
\end{eqnarray}
where the different moments are defined as
\begin{eqnarray}
    T^{n}=\int_{0}^{\infty}t^{n}p_{f}(t)dt
    \label{Tn}
\end{eqnarray}
and
\begin{eqnarray}
   p_{f}(s)=\int_0^\infty e^{-st}p_f(t)dt. 
   \label{pfs}
\end{eqnarray}
On the other hand, from the Laplace transform of \eqref{Q}, expanding $Q(s)$ for small $s$ yields 
\begin{eqnarray}
    Q(s)=Q(s=0)
+sQ'(s=0)+\frac{s^2}{2!}Q''(s=0)+\dots
\label{expq}
\end{eqnarray}
Taking into account the relation 
\begin{eqnarray}
    p_f(s)=1-sQ(s)
    \label{rela}
\end{eqnarray}
in the Laplace space and
equating the coefficients of the powers of $s$ we find the two first moments of the first passage PDF:
\begin{equation}
    T=Q(s=0),\quad T^{2}=-2Q'(s=0).
    \label{2m}
\end{equation}
From  \eqref{Q2} and \eqref{2m} the MFPT is then given by
\begin{equation}
    T(x_0)=Q(s=0)=\frac{1-e^{-|x_{0}-x_T|\sqrt{\frac{r}{D}}}}{r\int_{-\infty}^{\infty}e^{-|x_0-x_T|\sqrt{\frac{r}{D}}}f(x_0)dx_0},
\end{equation}
provided that $|x_{0}-x_T|=|x_{T}-x_0|$.
However, $x_0$ is a random variable drawn from $f(x_0)$ and the averaged mean first-passage time (AMFPT) over the possible values of $x_0$ is
\begin{equation}
\left\langle T\right\rangle _{x_{0}}=\int_{-\infty}^{\infty}T(x_0)f(x_{0})dx_{0}=\frac{1}{r}\left[\frac{1}{I(r,x_T)}-1\right]
    \label{ta1}
\end{equation}
with
\begin{equation}
  I(r,x_T)\equiv\int_{-\infty}^{\infty}e^{-|x_0-x_T|\sqrt{\frac{r}{D}}}f(x_0)dx_0,
  \label{I}
\end{equation}
which corresponds to the result found by Evans and Majumdar in \cite{EM2011}. Without losing generality, we will assume $x_T>0$  and will call by $\mathcal{D}$ the support of $f(x_0)$ from now on. Also, in the Appendix we prove (for sanity) that $I(r,x_T)<1$, so that $\left\langle T\right\rangle _{x_{0}}>0$.

\subsection{Effect of the resetting distribution on the first passage time PDF}
\label{universalpdf}

Here we are going to unveil some relevant properties of the survival probability derived in the previous section. If we average over the reset point PDF $f(x_0)$ (multiplying Eq. (\ref{Q2}) by $f(x_0)$ and integrating over $x_0$) we obtain


\begin{equation}
    \left\langle Q(s)\right\rangle _{x_{0}}= \frac{1-I(s+r,x_T)}{s+r I(s+r,x_T)}
    \label{aux1}
\end{equation}
where $I(r+s,x_{T})$ is given in Eq. \eqref{I} replacing
$r$ by $r+s$. We note that since the averaged survival probability only depends on the resetting distribution through the integral $I(s,x_T)$, a fixed form of this integral determines completely the survival properties of the target (and then the first passage time PDF). Putting this together with the result in (\ref{ta1}), we find that specifying the value of the AMFPT (which is equivalent to fix $I(s,x_T)$) the first-passage properties become univocally determined. This means that any resetting point PDF $f(x_0)$ leading to the same AMFPT will follow exactly the same first-passage statistics, revealing a somewhat universal behavior of the averaged first-passage distribution.

Going further, we could put expression (\ref{aux1}) in terms of $\langle T \rangle_{x_0}$ by inverting Eq. (\ref{ta1}). While a closed expression cannot be found in general, in the limit $s \ll r$ it is easy to see that one obtains
\begin{equation}
    \left\langle Q(s)\right\rangle _{x_{0}} \simeq \frac{\langle T \rangle _{x_0}}{1+s \langle T \rangle _{x_0}}
\end{equation}

Similarly, the first passage PDF (in the Laplace space) will read 
\begin{equation}
    \left\langle p_f(s) \right\rangle_{x_{0}} =1-s\left\langle Q(s)\right\rangle _{x_{0}}  \simeq \frac{1}{1+s \langle T \rangle _{x_0}}
\end{equation}

Thus, one finds that both the survival probability and the first-passage distribution averaged over the resetting points distribution are always exponential in the long time limit ($s \ll r$), and its dependence on $r$ and $f(x_0)$ is exclusively contained within the first moment $\langle T \rangle _{x_0}$.

\section{Symmetric resetting points PDF in an interval and $x_T\notin \mathcal{D}$}
In this section, we introduce some assumptions to get analytical results for the AMFPT. We consider that the resetting process is symmetric, this is, the probability of resetting to $x_0$ or $-x_0$ is the same. This means that $f(x_0)$ is an even function. 
Additionally, we assume $x_T\notin \mathcal{D}$, so from Eq. \eqref{I} we get
\begin{equation}
I(r,x_T)
=e^{-x_{T}\sqrt{\frac{r}{D}}}\int_{\mathcal{D}}e^{x_{0}\sqrt{\frac{r}{D}}}f(x_{0})dx_{0}.
\label{12}
\end{equation}
Then \eqref{ta1} becomes
\begin{equation}
    \left\langle T\right\rangle _{x_{0}}=\frac{1}{r}\left[\frac{e^{x_{T}\sqrt{\frac{r}{D}}}}{\int_{\mathcal{D}}f(x_0)\exp\left(x_0\sqrt{\frac{r}{D}}\right)dx_0}-1\right].
    \label{ta2}
\end{equation}

Using for instance $\mathcal{D}=[-L,L]$, then
\begin{equation}
\left\langle T\right\rangle _{x_{0}}(r,L)=\frac{1}{r}\left[\frac{e^{x_{T}\sqrt{\frac{r}{D}}}}{\int_{-L}^{L}f(x_0)e^{-x_0\sqrt{\frac{r}{D}}}dx_0}-1\right].
    \label{ta3}
\end{equation}
In the limit $r\to 0^+$ the integral in Eq. \eqref{ta3} tends to 1 so that $\left\langle T\right\rangle _{x_{0}}\sim r^{-1}$; this means that the AMFPT diverges to $+\infty$in this limit. On the other side, for the limit $r\to\infty$ we can find a lower bound for $\left\langle T\right\rangle _{x_{0}}$, so if we prove that the lower bound diverges as $r\to\infty$, then $\left\langle T\right\rangle _{x_{0}}$ will do. If we consider then
\begin{eqnarray*}
    \int_{-L}^{L}f(x_0)e^{-x_0\sqrt{\frac{r}{D}}}dx_0&<&\underset{x_0}{\max}\left(e^{-x_0\sqrt{\frac{r}{D}}}\right)\int_{-L}^{L}f(x_0)dx_0\\
    &=&e^{L\sqrt{\frac{r}{D}}}
\end{eqnarray*}
then we see that
$$
\left\langle T\right\rangle _{x_{0}}(r,L)>\frac{1}{r}\left[e^{(x_{T}-L)\sqrt{\frac{r}{D}}}-1\right].
$$
is satisfied. Since $x_T>L$, in the limit $r\to\infty$ the lower bound tends to infinity and thus $\left\langle T\right\rangle _{x_{0}}\to\infty$. Therefore, we conclude that there exists a value of $r$ which optimizes the AMFPT when the resetting points are symmetrically distributed in a finite interval. This is the first main result of this work. Next, we study how the AMFPT depends on $L$ and find the optimal reset rate $r$ in the cases where it exists; to this end, we need to consider specific choices for $f(x_0)$.

\subsection{Example: Uniformly distributed reset points in an interval}
\begin{figure}[htbp]
    \includegraphics[width=0.8\hsize]{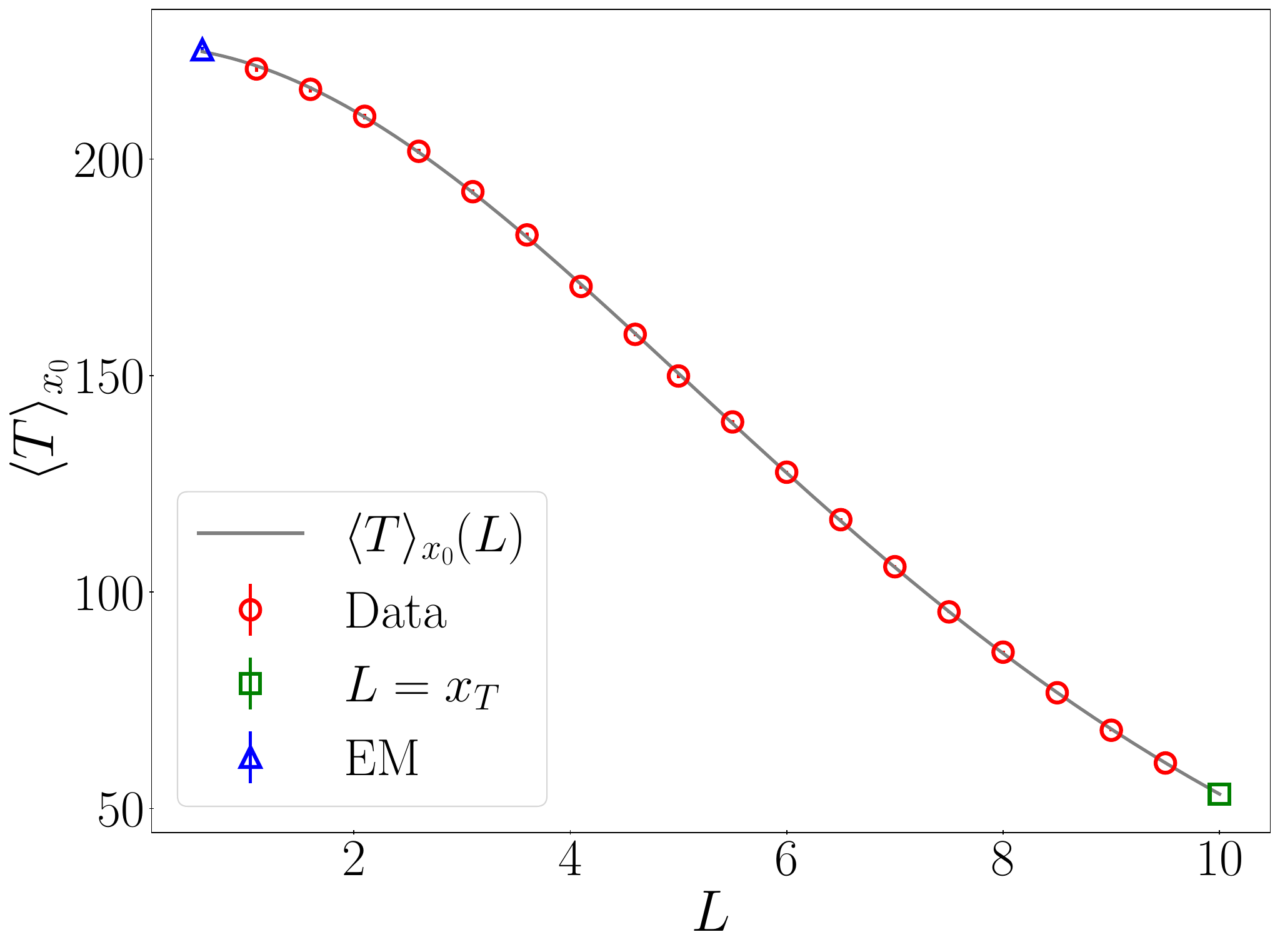}
    \caption{AMFPT for uniformly distributed reset points PDF in an interval $\left[-L,L\right]$ for $x_T\notin \mathcal{D}$. The solid line represents Eq. \eqref{Ts}. The data points are obtained from numerical simulations. In blue (triangle) we present the Evans and Majumdar result, Eq. \eqref{EM}, and in green (square) the case $L \to x_T$, Eq. \eqref{eq:MFPT_uniform}. The parameters used for the numerical simulations are $D=1$, $r=0.1$, and $x_T =10$. We have used $N=10^5$ trajectories.}
    \label{fig:fig1}
\end{figure}

We assume first that the reset points $x_0$ are uniformly distributed in the interval $[-L,L]$. That is, $f(x_0)=1/2L$ if $x_0\in [-L,L]$, and zero otherwise. Therefore, from \eqref{I}
\begin{eqnarray}
 I(r,x_T,L)=\frac{1}{L}\sqrt{\frac{D}{r}}e^{-x_{T}\sqrt{\frac{r}{D}}}\sinh\left(L\sqrt{\frac{r}{D}}\right),  
 \label{Iud}
\end{eqnarray}

so \eqref{ta2} reduces to
\begin{equation}
    \left\langle T\right\rangle _{x_{0}}(r,L)=\frac{1}{r}\left[L\sqrt{\frac{r}{D}}\frac{e^{x_{T}\sqrt{\frac{r}{D}}}}{\sinh\left(L\sqrt{{\frac{r}{D}}}\right)}-1\right].
    \label{Ts}
\end{equation}
When $L\to 0$, the PDF of resetting points tends to a Dirac delta function and Eq. \eqref{Ts} reduces to the Evans and Majumdar (EM) result in \cite{EM2011prl}, that is, when the searcher resets to $x_0=0$
\begin{equation}
    \left\langle T\right\rangle _{x_{0}}(r,L\to 0)=\frac{1}{r}\left(e^{x_{T}\sqrt{\frac{r}{D}}}-1\right).
    \label{EM}
\end{equation}
On the other hand, when $x_T \to L$,  Eq. \eqref{Ts} becomes
\begin{equation}
    \left\langle T\right\rangle _{x_{0}}(r,L=x_T)=\frac{1}{r}\left[L\sqrt{\frac{r}{D}}\frac{2}{1-e^{-2L\sqrt{\frac{r}{D}}}}-1\right].
    \label{eq:MFPT_uniform}
\end{equation}
In Figure \ref{fig:fig1} we plot $\left\langle T\right\rangle _{x_{0}}(r,L)$ with respect to $L$ for fixed values of the rest of parameters. The agreement between the simulations (circles) and 
\eqref{Ts} is excellent. The particular cases \eqref{EM} and \eqref{eq:MFPT_uniform} are also shown in the figure (blue and green symbols). As can be seen, the AMFPT decreases monotonically with the width of the resetting points interval, i.e., resetting to a point (this is
the case $L = 0$) is less effective (the AMFPT is higher)
than resetting to uniformly distributed points. 

\begin{figure}[htbp]
    \includegraphics[width=0.8\hsize]{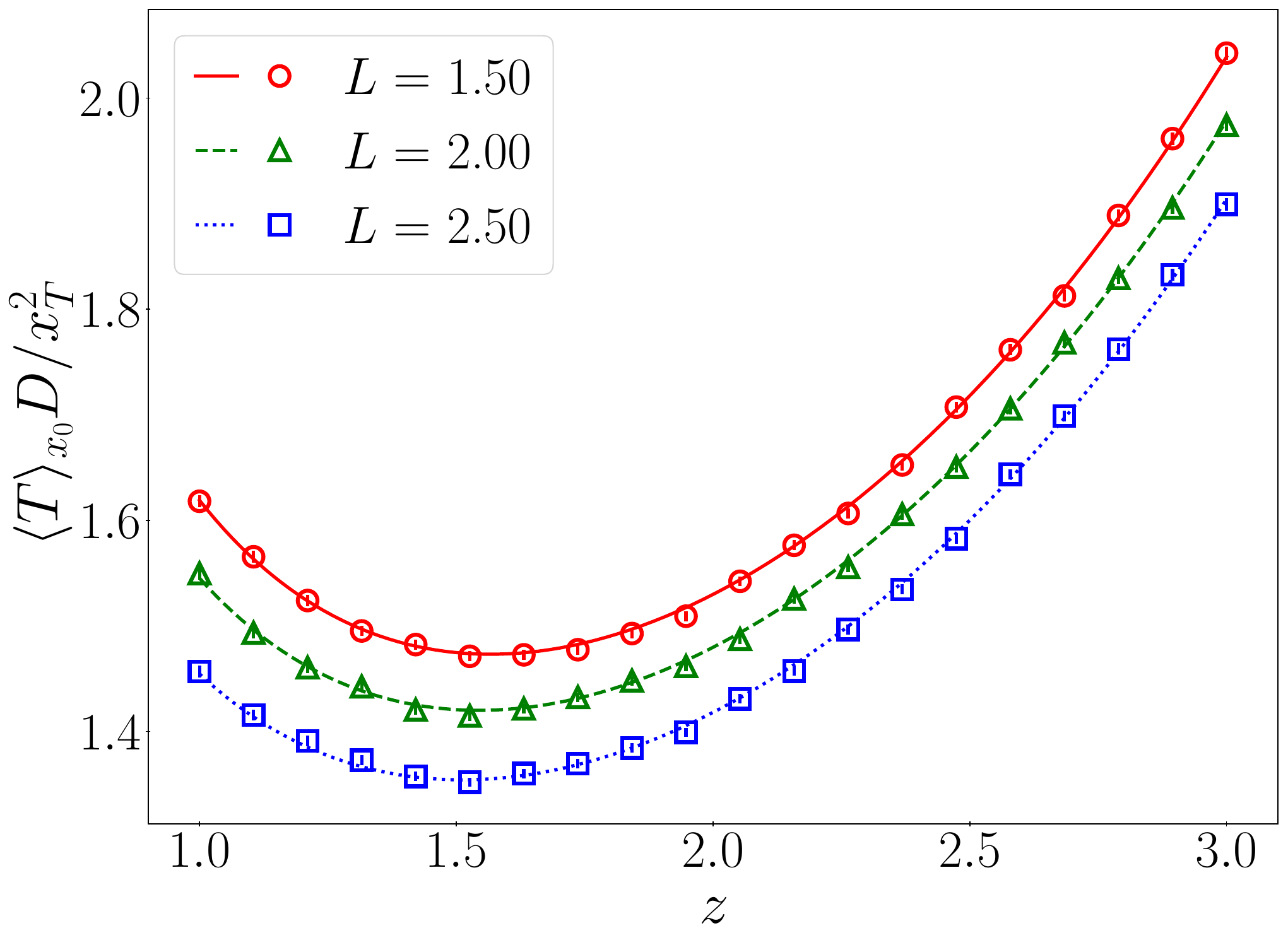}
    \caption{ $\frac{D}{x_T^{2}}\left\langle T\right\rangle _{x_{0}}$ as function of the parameter $z=x_T\sqrt{r/D}$ for three values of $L$, $L = 1.50$ (red), $L = 2.00$ (green) and $L = 2.50$ (blue). The lines represent Eq. \eqref{Tad18} while the points (circles, triangles and squares) are data from numerical simulations. The parameters used are $D=1$, $r=0.1$, and $x_T$ is computed given a fixed value of $z$. For the simulations we have used $N=10^5$ trajectories.}
    \label{fig:fig2}
\end{figure}

We are also interested in finding the optimal reset rate that minimizes the AMFPT, namely $r_{opt}$. Let us introduce the following dimensionless quantities
\begin{eqnarray}
    \epsilon = L/x_T,\quad z=x_T\sqrt{r/D}.
    \label{def}
\end{eqnarray}
Since $x_T>L$, then $0<\epsilon<1$ and from Eq. \eqref{Ts}
\begin{eqnarray}
    \frac{D}{x_T^{2}}\left\langle T\right\rangle _{x_{0}}=\frac{\epsilon e^{z}}{z\sinh(\epsilon z)}-\frac{1}{z^{2}}.
    \label{Tad18}
\end{eqnarray}
Note that the case $\epsilon =0$ reduces to the EM case. In Figure \ref{fig:fig2} we observe the existence of an optimal reset rate, $r_{opt}$, for different values of $L$ and compare the result given in Eq. \eqref{Tad18} with numerical simulations. In agreement with Figure \ref{fig:fig1} we observe that curves with large $L$ lead to lower values of the AMFPT.

Computing $d\left\langle T\right\rangle _{x_{0}}/dz=0$ we obtain a transcendent equation for $z$ whose solution, namely $z_{opt}$, allows to compute the optimal reset rate as  
\begin{eqnarray}
   r_{opt}=\frac{D}{x_T^2}z_{opt}^2. 
   \label{ropt}
\end{eqnarray}

\begin{figure}[htbp]
    \includegraphics[width=0.8\hsize]{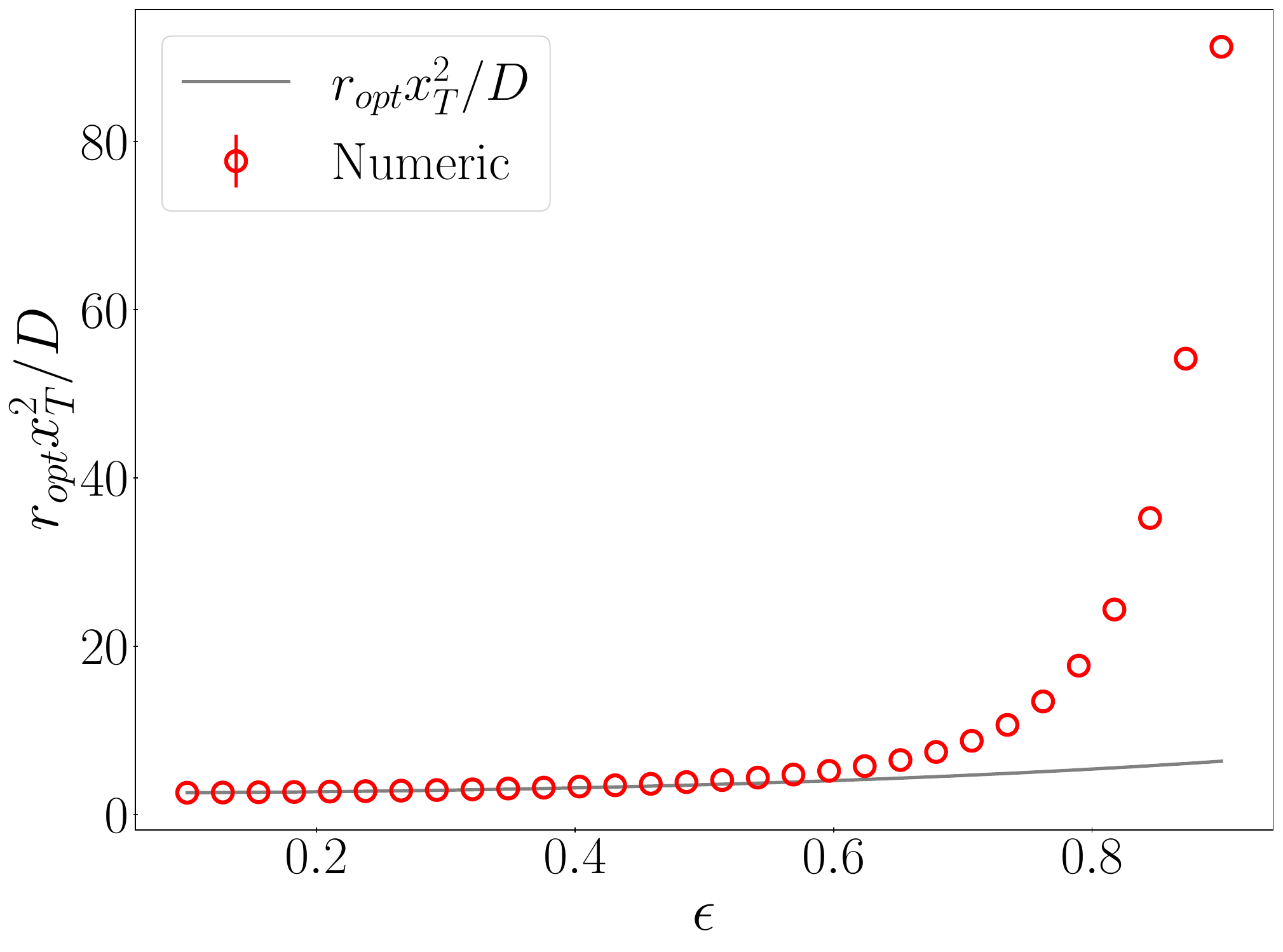}
    \caption{$r_{opt}x_T^2/D$ vs $\epsilon$. The solid line is computed using Eq. \eqref{ro1} while the data points are obtained by numerically computing $d\left\langle T\right\rangle _{x_{0}}/dz=0$.}
    \label{fig:fig3}
\end{figure}

We can obtain approximated analytic results for the optimal reset rate by introducing the perturbative expansion $z=z_0+z_1\epsilon+z_2\epsilon^2+...$ into the equation for $d\left\langle T\right\rangle _{x_{0}}/dz=0$. Equating the coefficients of the powers of $\epsilon$ we can calculate $z_0$, $z_1$, ... The $O(0)$ coefficient is $2+(z_0-2)e^{z_0}=0$ whose solution is $z_0=1.5936$ and corresponds to the value found in \cite{EM2011prl}. The next order, $O(1)$ gives $z_1=0$ while the second order gives
$$
z_{2}=-\frac{z_{0}^{3}\left(z_{0}e^{z_{0}}-4e^{z_{0}}+4\right)}{6\left(z_{0}^{2}e^{z_{0}}+4-4e^{z_{0}}+z_{0}e^{z_{0}}\right)}=1.1363.
$$
Finally, from \eqref{def}
\begin{eqnarray}
    r_{opt}=2.5396\frac{D}{x_{T}^{2}}\left[1+0.713\left(\frac{L}{x_{T}}\right)^{2}+...\right]^2
    \label{ro1}
\end{eqnarray}
where the prefactor corresponds to the value of the optimal reset rate found by EM, and we have found the lowest order correction term due to the uniform resetting in the interval $[-L,L]$. We compare in Figure \ref{fig:fig3} the numerical solution to the equation $d\left\langle T\right\rangle _{x_{0}}/dz=0$ (circles) with the approximation provided by Eq. \eqref{ro1} (curve). We note that when the target is far from the resetting interval, $x_ T\gg L$ that is $\epsilon \to 0$, the optimal situation takes place when the reset rate is low and so the searcher has enough time to reach the target. Then, as $\epsilon \to 0$ we expect $r_{opt}\to 0$. This is observed in Figure \ref{fig:fig3}. Contrarily, as $x_T\to L$, i.e., $\epsilon \to 1$, to increase the chance of reaching the target, the optimal strategy is to avoid large excursions which implies increasing the reset rate. This is also observed in Figure \ref{fig:fig3} where $r_{opt}$ increases with $\epsilon$.  

\section{Symmetric resetting points PDF and $x_T\in \mathcal{D}$}
In this Section we consider the case $x_T\in \mathcal{D}$ when the resetting points are symmetrically distributed in an interval (i.e., $f(x_0)$ has finite support) or in the real line ($f(x_0)$ has infinite support).

\subsection{Finite support}
\begin{figure}[htbp]
    \includegraphics[width=0.8\hsize]{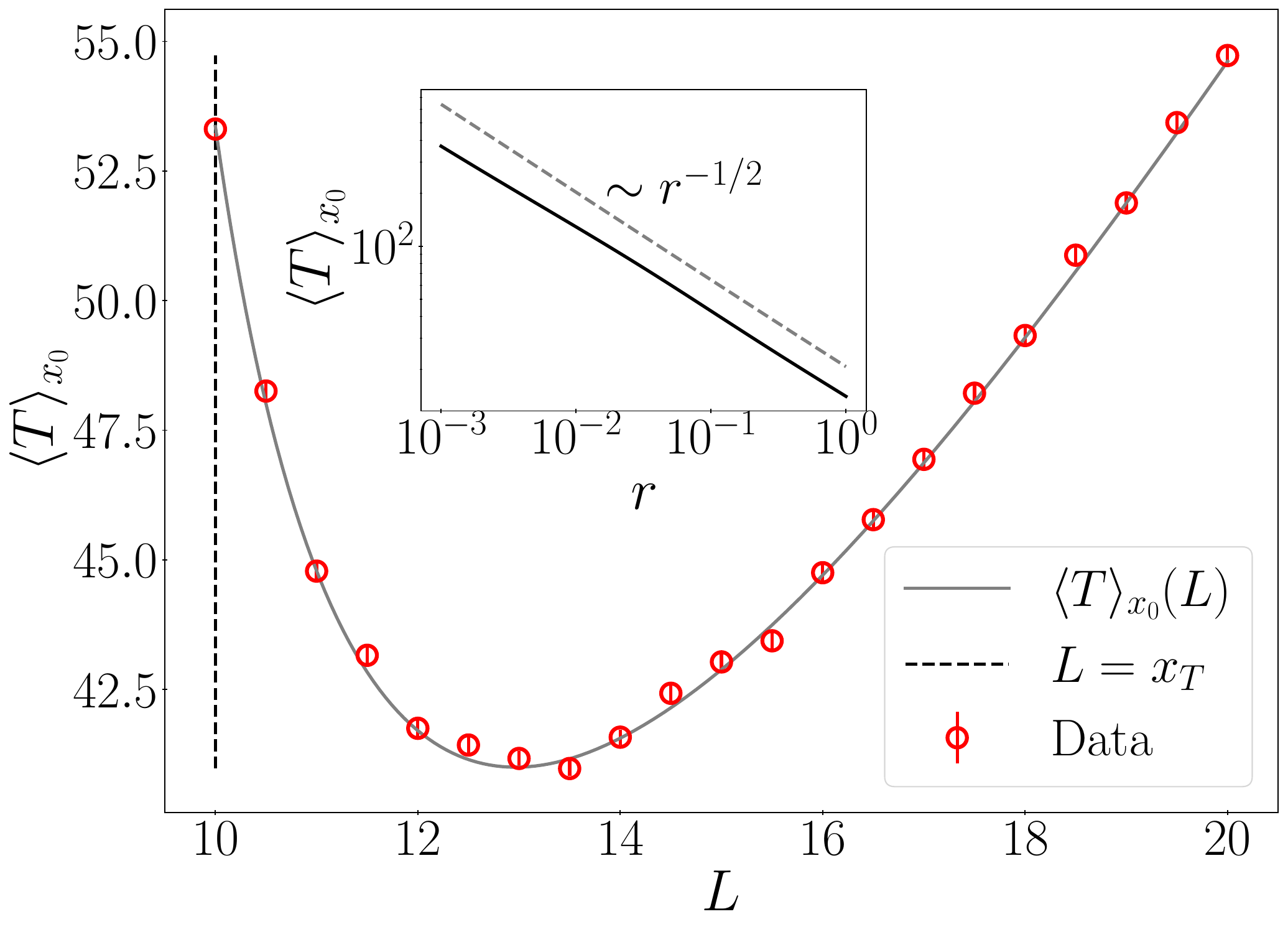}
    \caption{AMFPT for uniformly distributed reset points PDF in an interval $\left[-L,L\right]$ for $x_T\in \mathcal{D}$. The solid line represents Eq. \eqref{ta5}. The data points are obtained from numerical simulations. The parameters used for the numerical simulations are $D=1$, $r=0.1$, and $x_T =10$. We have used $N=10^5$ trajectories. In the inset, the solid line represents the AMFPT as a function of the resetting rate $r$ for $L=15$ from Eq. \eqref{ta5}. The dashed line shows the predicted scaling as $r^{-1/2}$.}
    \label{fig:fig4}
\end{figure}

Let us assume first  $\mathcal{D}=[-L,L]$, and so $-L<x_T<L$; then the value of $L$ ranges from $x_T$ to $\infty$. From the first equality in \eqref{cas2} and considering that $f(x_0)$ is an even function we get
\begin{eqnarray}
I(r,x_{T},L)&=&e^{-x_{T}\sqrt{\frac{r}{D}}}\int_{-L}^{x_{T}}f(x_{0})e^{x_{0}\sqrt{\frac{r}{D}}}dx_{0}\nonumber\\
&+&e^{x_{T}\sqrt{\frac{r}{D}}}\int_{x_{T}}^{L}f(x_{0})e^{-x_{0}\sqrt{\frac{r}{D}}}dx_{0}.
    \label{I1}
\end{eqnarray}
If the resetting point PDF $f(x_0)$ is uniform in $[-L,L]$, then
\begin{eqnarray}
  I(r,x_{T},L)=\frac{1}{L}\sqrt{\frac{D}{r}}\left[1-e^{-L\sqrt{\frac{r}{D}}}\cosh\left(x_{T}\sqrt{\frac{r}{D}}\right)\right]  
  \label{I4}
\end{eqnarray}
and from \eqref{ta1} the AMFPT reads 
\begin{eqnarray}
    \left\langle T\right\rangle _{x_{0}}(r,L)=\frac{1}{r}\left[\frac{L\sqrt{\frac{r}{D}}}{1-e^{-L\sqrt{\frac{r}{D}}}\cosh\left(x_{T}\sqrt{\frac{r}{D}}\right)}-1\right].
    \label{ta5}
\end{eqnarray}
In Figure \ref{fig:fig4} we check Eq. \eqref{ta5} with numerical simulations, for both the behavior of the AMFPT with respect to $r$ and with respect to $L$, showing an excellent agreement. 

To prove that there exists always an optimal $L$ which minimizes the AMFPT we study its behavior both for $L\to x_T$ and for large $L$. On one hand, it can be shown that $\left[\partial\left\langle T\right\rangle _{x_{0}}(L)/\partial L\right]_{L=x_{T}}<0$. On the other hand, from Eq. \eqref{ta5} it is easy to see that $\left\langle T\right\rangle _{x_{0}}(L)\sim L$ as $L\to \infty$.
The optimal resetting domain width, say $L_{opt}$, can be found from \eqref{ta5} by solving the equation $\partial \left\langle T\right\rangle _{x_{0}}/\partial L=0$ to get
\begin{eqnarray}
    L_{opt}=\sqrt{\frac{D}{r}}\left[-1-W_{-1}\left(-\frac{1}{e\cosh\left(x_{T}\sqrt{\frac{r}{D}}\right)}\right)\right],
    \label{lopt}
\end{eqnarray}
where $W_{-1}(\cdot)$ is the lower branch of the Lambert function. For example, for the data used in Fig. \ref{fig:fig4} from \eqref{lopt} we find $L_{opt}=12.97$, which seems to agree with the location of the minimum observed in Fig \ref{fig:fig4}. At $L=x_T$ the target is located at the right limit on the resetting interval and the AMFPT reaches the value
$$
\left\langle T\right\rangle _{x_{0}}(r,L=x_{T})=\frac{2x_{T}/\sqrt{rD}}{1-e^{-2x_{T}\sqrt{\frac{r}{D}}}}.
$$
As $L$ departs from $x_T$ and increases, to reach the target is more likely and the AMFPT decreases. On the other side, if $L$ is large enough then an increase of $L$ yields a larger AMFPT. Then, there must be an optimal $L$ which minimizes the AMFPT as shown in Figure \ref{fig:fig4}.

Finally, we are also interested in finding if there exists an optimal reset rate as in the previous example. In the limit where $r$ approaches to 0 from the left we find the behaviour $\left\langle T\right\rangle _{x_{0}}\sim r^{-1/2}$, so that \eqref{ta5} diverges in this limit. In the limit $r\to\infty$ then it is shown that $\left\langle T\right\rangle _{x_{0}}\sim r^{-1/2}$ as well. Altogether, no optimal reset rate exists and the scaling behavior $\left\langle T\right\rangle _{x_{0}}\sim r^{-1/2}$ is observed both for small and large $r$.  This is shown in the inset of Figure \ref{fig:fig4} where the scaling $r^{-1/2}$ is observed within some orders of magnitude.

\subsection{Infinite support}
\begin{figure}[htbp]
    \includegraphics[width=0.8\hsize]{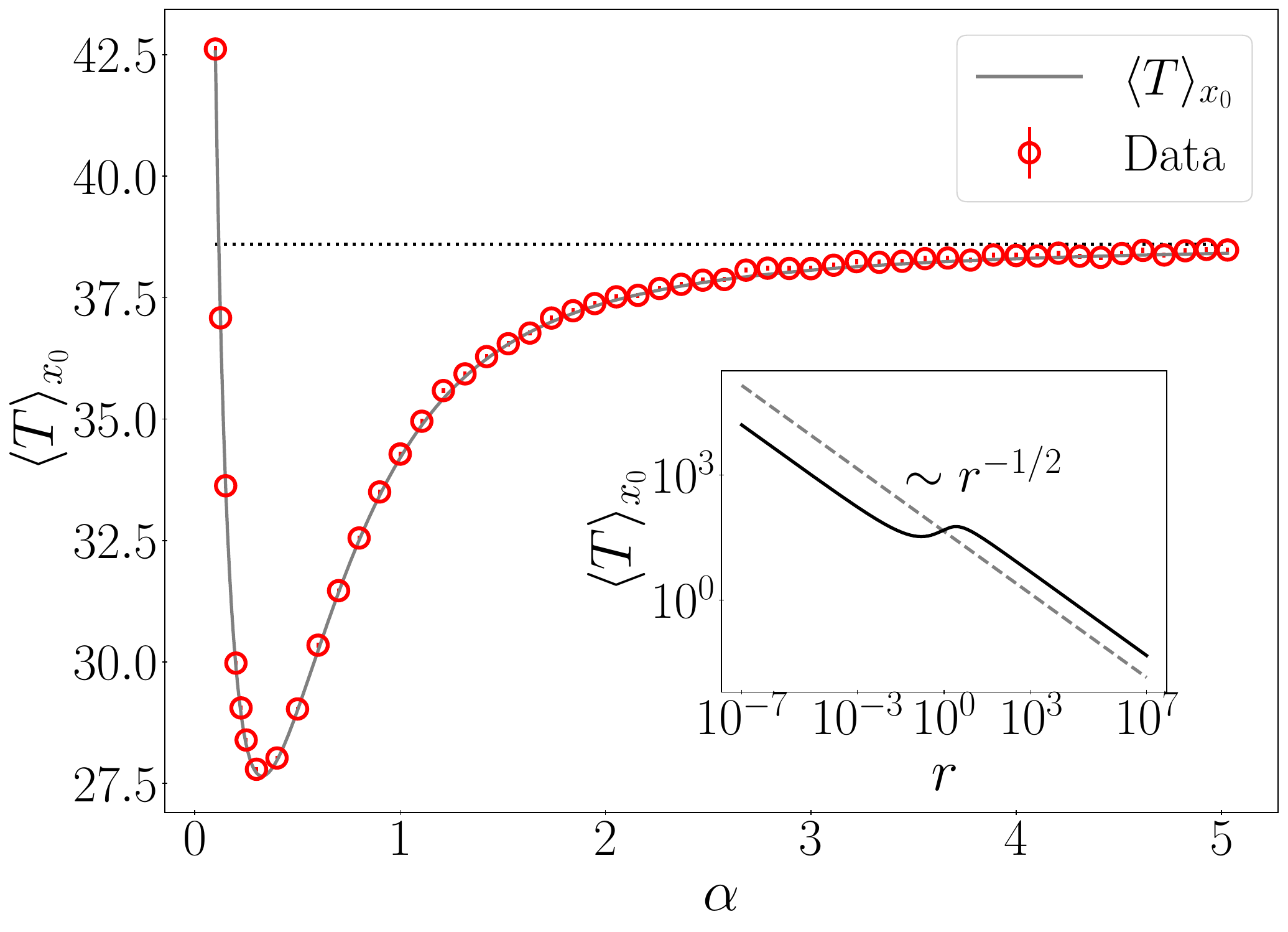}
    \caption{AMFPT for exponentially distributed reset points as a function of $\alpha$. The solid line represents Eq. \eqref{ta7}, the data points are obtained from numerical simulations and the dotted line represents the value of $\langle T \rangle_{x_0}$ for the EM case (Eq. \eqref{EM}). The parameters used for the simulations are $D=1$, $r=0.1$, and $x_T =5$. We have used $N=5\cdot 10^5$ trajectories. In the inset, the solid line represents the AMFPT as a function of the resetting rate $r$ for $\alpha=1$ from Eq. \eqref{ta7}. The dashed line shows the decaying as $r^{-1/2}$. Both axes are on a logarithmic scale.}
    \label{fig:fig5}
\end{figure}

The second situation consists in considering $\mathcal{D}=\mathbb{R}$. Let us assume for example that the resetting point is exponentially distributed. If we consider
\begin{eqnarray}
 f(x_0)=\frac{\alpha}{2}e^{-\alpha |x_0|},\quad \alpha>0  
 \label{exp}
\end{eqnarray}
into Eq. \eqref{ta1}, we get
$$
I(r,x_{T},L)=\frac{\alpha}{\alpha^{2}-\frac{r}{D}}\left(\alpha e^{-x_{T}\sqrt{\frac{r}{D}}}-\sqrt{\frac{r}{D}}e^{-x_{T}\alpha}\right).
$$
The AMFPT finally reads from  \eqref{ta1}
\begin{eqnarray}
\left\langle T\right\rangle _{x_{0}}(r,\alpha)=\frac{1}{r}\left(\frac{\alpha^{2}-\frac{r}{D}}{\alpha^{2}e^{-x_{T}\sqrt{\frac{r}{D}}}-\alpha\sqrt{\frac{r}{D}}e^{-x_{T}\alpha}}-1\right).
    \label{ta7}
\end{eqnarray}
Taking the limit $r\to 0^+$ to Eq. \eqref{ta7} we find $\left\langle T\right\rangle _{x_{0}}\sim r^{-1/2}$. Taking the limit $r\to \infty$ to \eqref{ta7} one has $\left\langle T\right\rangle _{x_{0}}\sim r^{-1/2}$ as well, and then no optimal reset rate exists. This scaling can be seen in the inset of Figure \ref{fig:fig5}.

From \eqref{ta7} it results that $\left\langle T\right\rangle _{x_{0}}\sim \alpha^{-1}$ as $\alpha \to 0^+$, while in the limit $\alpha\to \infty$ the exponential function in Eq. \eqref{exp}
tends to a delta function so $\left\langle T\right\rangle _{x_{0}}$ tends to a constant value corresponding to the EM result in Eq. \eqref{EM}. For example, considering the specific case shown in Fig. \ref{fig:fig5} and using \eqref{EM}, the AMFPT in the limit of large $\alpha$ is 38.6.

\section{resetting to two points}
Next, we consider the situation in which the resetting process is not symmetric. Recall that \eqref{ta2} holds only for symmetric $f(x_0)$, so here we need to make use of the general expression \eqref{I} to compute the AMFPT.
Let us consider for the sake of simplicity the resetting points PDF given by
\begin{equation}
  f(x_0)=p\delta (x_0-L)+(1-p)\delta (x_0+L)  ,
  \label{2d}
\end{equation}
which means that the searcher either resets to the point $x_0=+L$ with probability $p$, or to the point $x_0=-L$ with probability $1-p$.

If $x_T>L$ then, from \eqref{ta1} and \eqref{2d}
\begin{equation}
    \left\langle T\right\rangle _{x_{0}}(r,L)=\frac{1}{r}\left[\frac{e^{x_{T}\sqrt{\frac{r}{D}}}}{(1-p)e^{-L\sqrt{\frac{r}{D}}}+pe^{L\sqrt{\frac{r}{D}}}}-1\right].
    \label{t27}
\end{equation}
It can be shown that $\left\langle T\right\rangle _{x_{0}}$ approaches  EM result \eqref{EM} in the limit $L\to 0$. For $L>0$, $\left\langle T\right\rangle _{x_{0}}$ reaches a maximum value at 
$$
L=L_{c}=\sqrt{\frac{D}{r}}\ln\left(\frac{\sqrt{p(1-p)}}{p}\right)
$$
provided that $0<p<1/2$. For $1/2<p<1$, $L_c$ does not exist and the maximum of  $\left\langle T\right\rangle _{x_{0}}$ is attained at $L=0$. However, when $0<p<1/2$ it can be shown that 
$\left\langle T\right\rangle _{x_{0}}(L=0)<\left\langle T\right\rangle _{x_{0}}(L)$ for $0<L<L^*$ and $\left\langle T\right\rangle _{x_{0}}(L=0)>\left\langle T\right\rangle _{x_{0}}(L)$ for $L^*<L<x_T$, where we have defined
\begin{equation}
L^*=\sqrt{\frac{D}{r}}\ln\left(\frac{1-p}{p}\right).
\label{Lstar}
\end{equation}
\begin{figure}[htbp]
    \includegraphics[width=\hsize]{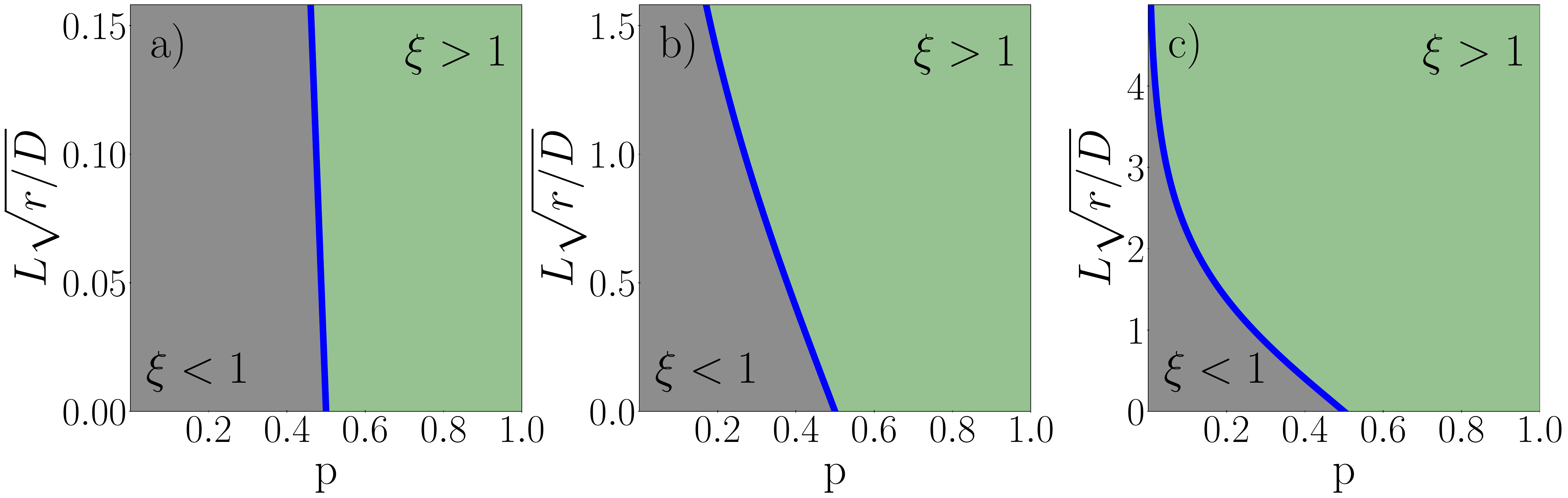}
    \caption{Diagram $L\sqrt{r/D}$ vs $p$ for asymmetric resetting points in an interval with $x_T>L$. The parameter $\xi$ is defined as $\xi = \left\langle T\right\rangle _{x_{0}}(L=0)/\left\langle T\right\rangle _{x_{0}}(L)$. We have obtained the value of $\xi$ numerically from Eq.\eqref{EM} and Eq.\eqref{t27}. The green region ($\xi>1$) corresponds to the case where the asymmetric resetting strategy is more efficient, while in the grey region the most efficient strategy is to reset to $x_0=0$. The blue line, computed through Eq. \eqref{Lstar}, represents $L^*\sqrt{r/D}$ vs $p$. Panel a) is for a reset rate $r=0.01$, panel b) for $r=0.1$ and panel c) for $r=1$.}
    \label{fig:fig6}
\end{figure}

In Figure \ref{fig:fig6} we summarize these results graphically.  For $x_T<L$ and using \eqref{ta1} and \eqref{2d}
\begin{equation}
     \left\langle T\right\rangle _{x_{0}}(r,L)=\frac{1}{r}\left[\frac{e^{L\sqrt{\frac{r}{D}}}}{(1-p)e^{-x_T\sqrt{\frac{r}{D}}}+pe^{x_T\sqrt{\frac{r}{D}}}}-1\right] 
     \label{T4}
\end{equation}
i.e., the AMFPT is monotonically increasing with $L$ for $L\in(x_T,\infty)$. Note that this expression is the same as Eq. \eqref{t27} but exchanging $x_T$ and $L$.

To study the optimal reset rate that minimizes the AMFPT we proceed as above. It is easy to show that in the limit $r\to 0^+$ the AMFPTs given by \eqref{t27} and \eqref{T4} go as $r^{-1}$ as in the previous examples. Hence, the AMFPT diverges when $r$ approaches to 0. In the limit $r\to \infty$ from \eqref{t27} we have $\left\langle T\right\rangle _{x_{0}}\sim\frac{e^{(x_T-L)\sqrt{r/D}}}{r}$, and this goes to infinity provided that $x_T>L$. For $x_T<L$, using \eqref{T4} we have $\left\langle T\right\rangle _{x_{0}}\sim\frac{e^{(L-x_{T})\sqrt{
r/D}}}{r}$ as $r\to\infty$, so the existence of an optimal reset rate is proven if $x_T\neq L$. However, for the specific case $x_T=L$ we have $\left\langle T\right\rangle _{x_{0}}\sim r^{-1}$ and no optimal reset rate exists; the minimum value of the AMFPT is zero and it is attained as $r\to \infty$. This means that the optimal reset rate diverges as $L$ tends to $x_T$. By proposing the ansatz 
\begin{eqnarray}
r_{opt}\sim |\epsilon -1|^{-\beta},\quad \textrm{as}\quad \epsilon\to 1.
    \label{guess}
\end{eqnarray}
 one can determine numerically the exponent $\beta$ has to be determined. 
 
\begin{figure}[htbp]
    \includegraphics[width=\hsize]{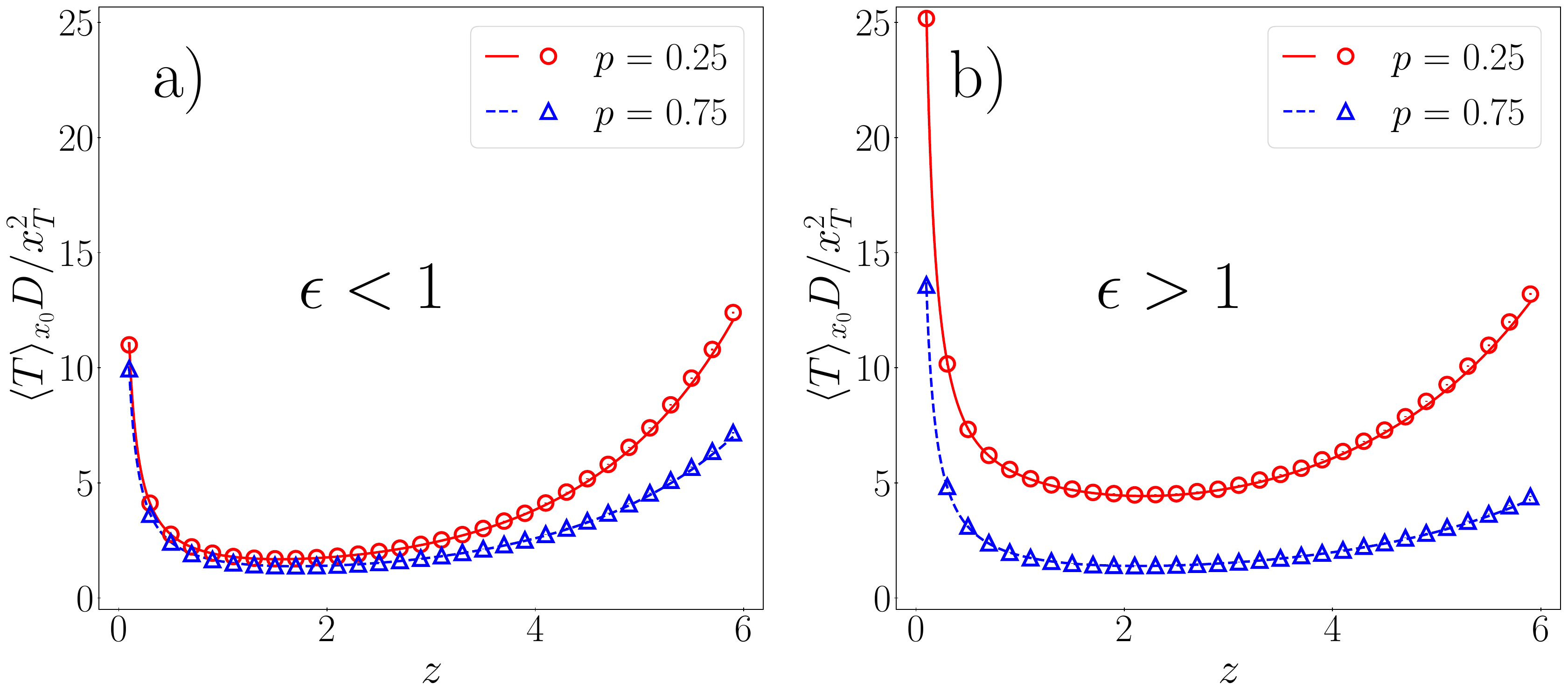}
    \caption{ $\frac{D}{x_T^{2}}\left\langle T\right\rangle _{x_{0}}$ as function of the parameter $z=x_T\sqrt{r/D}$ for $p=0.25$ (red) and $p=0.75$ (blue). The lines represent Eq. \eqref{sev4} while the points (circles and triangles) are data from numerical simulations. The parameters used for obtaining the data are $D=1$, $x_T=2$ and $r$ is computed given a fixed value of $z$. In panel a) $\epsilon=0.2$ thus $L=0.2$. In panel b) $\epsilon=1.8$ thus $L=3.6$. For the simulations we have used $N=10^5$ trajectories.}
    \label{fig:fig7}
\end{figure}

\begin{figure}[htbp]
    \includegraphics[width=\hsize]{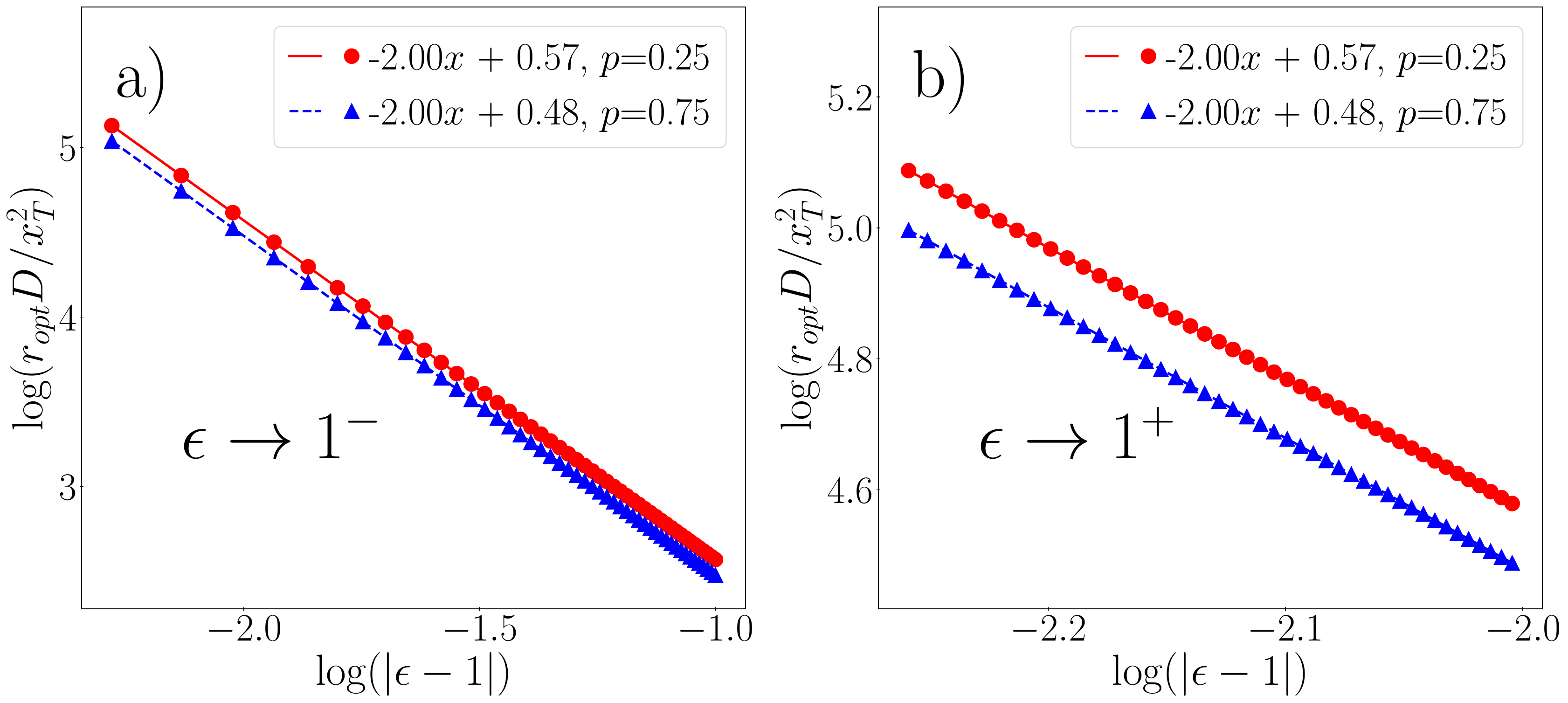}
    \caption{$r_{opt}(\epsilon)$ for $\epsilon \to 1$ for $p=0.25$ (red circles) and $p=0.75$ (blue triangles). The dots are obtained by numerically minimizing Eq. \eqref{sev4} and the lines are linear regressions to the data, for each curve the correlation coefficient $R^2=0.9999$. In panel a) $\epsilon\to1^-$. In panel b)  $\epsilon\to1^+$.}
    \label{fig:fig8}
\end{figure}

 We proceed numerically as in the previous section. Considering \eqref{def} into \eqref{t27} and \eqref{T4} we find
\begin{eqnarray}
   \frac{D}{x_{T}^{2}}\left\langle T\right\rangle _{x_{0}}=\left\{ \begin{array}{cc}
\frac{1}{z^{2}}\left[\frac{e^{z}}{(1-p)e^{-\epsilon z}+pe^{\epsilon z}}-1\right], & 0<\epsilon<1\\
\frac{1}{z^{2}}\left[\frac{e^{\epsilon z}}{(1-p)e^{-z}+pe^{z}}-1\right], & \epsilon>1.
\end{array}\right.
\label{sev4}
\end{eqnarray}
In Figure \ref{fig:fig7} we compare this result against numerical simulations. The agreement is excellent for both cases $\epsilon<1$ and $\epsilon>1$. In agreement with our reasoning, in both cases there is an optimal reset rate that minimizes the AMFPT. In the case $\epsilon \to 1$ we observe (see Figure \ref{fig:fig8}) that 
\begin{eqnarray}
    r_{opt}\sim \frac{D/x_T^2}{|\epsilon -1|^2}
    \label{scal}
\end{eqnarray}
so that the exponent $\beta =2$. To show this analytically  we make use of Eq. \eqref{ropt} and \eqref{guess} and write $z_{opt}\sim |\delta|^{-\beta}$, where $\delta = \epsilon\pm 1$. It is important to note that there will appear terms proportional to $z$ and to $z\delta$ and in the limit $\epsilon\to 1^{\pm}$, i.e., $\delta\to 0$ we have $z\to \infty$ and $z\gg z\delta$. 

Let us deal with the cases $\epsilon >1$ and $\epsilon<1$ separately. By taking the derivative of \eqref{sev4} respect to $z$ for $\epsilon<1$ we find that the $z_{opt}$ is solution to
\begin{eqnarray}
    p(\delta z-2)e^{4z}e^{-3z\delta}&+&(1-p)(-2+2z-\delta z)e^{2z}e^{-z\delta}\nonumber\\
    &=&2\left(pe^{2z}e^{-2z\delta}+1-p\right)^{2}.
    \label{eqm}
\end{eqnarray}
It is not difficult to see that the first term on the left-hand side (lhs) of Eq. \eqref{eqm} is much higher than the second term in the limit $\delta\to 0$ ($z\to\infty$) where $\delta = 1-\epsilon$. The term of the right-hand side (rhs) of Eq. \eqref{eqm} approaches $2p^2e^{4z}$. Eq. \eqref{eqm} reduces then to $\delta z_{opt}\sim 2p(1+p)$, so that $\delta^{1-\beta/2}\sim O(\delta^0)$, leading to $\beta=2$. 

We can proceed analogously for $\epsilon>1$. Taking the derivative of \eqref{sev4} with respect to $z$ we find 
\begin{eqnarray}
(1&-&p)\left(2-2z-z\delta\right)e^{2z}e^{z\delta}+p(2-\delta z)e^{4z}e^{\delta z}\nonumber\\
&=&2\left(pe^{2z}+1-p\right)^{2}
    \label{me2}
\end{eqnarray}
for $\epsilon >1$ where now $\delta =\epsilon -1$. In the limit $\delta\to 0$ we see that the second term of the lhs of Eq. \eqref{me2} is much higher than the first one. So that, $p(2-\delta z_{opt})\sim 2p^2$, i.e., $\delta z_{opt}\sim 2(1-p)$, which leads to $\delta^{1-\beta/2}\sim O(\delta^0)$ and then to $\beta=2$ again.

\section{The equivalent resetting point}
\begin{figure}[htbp]
    \includegraphics[width=0.8\hsize]{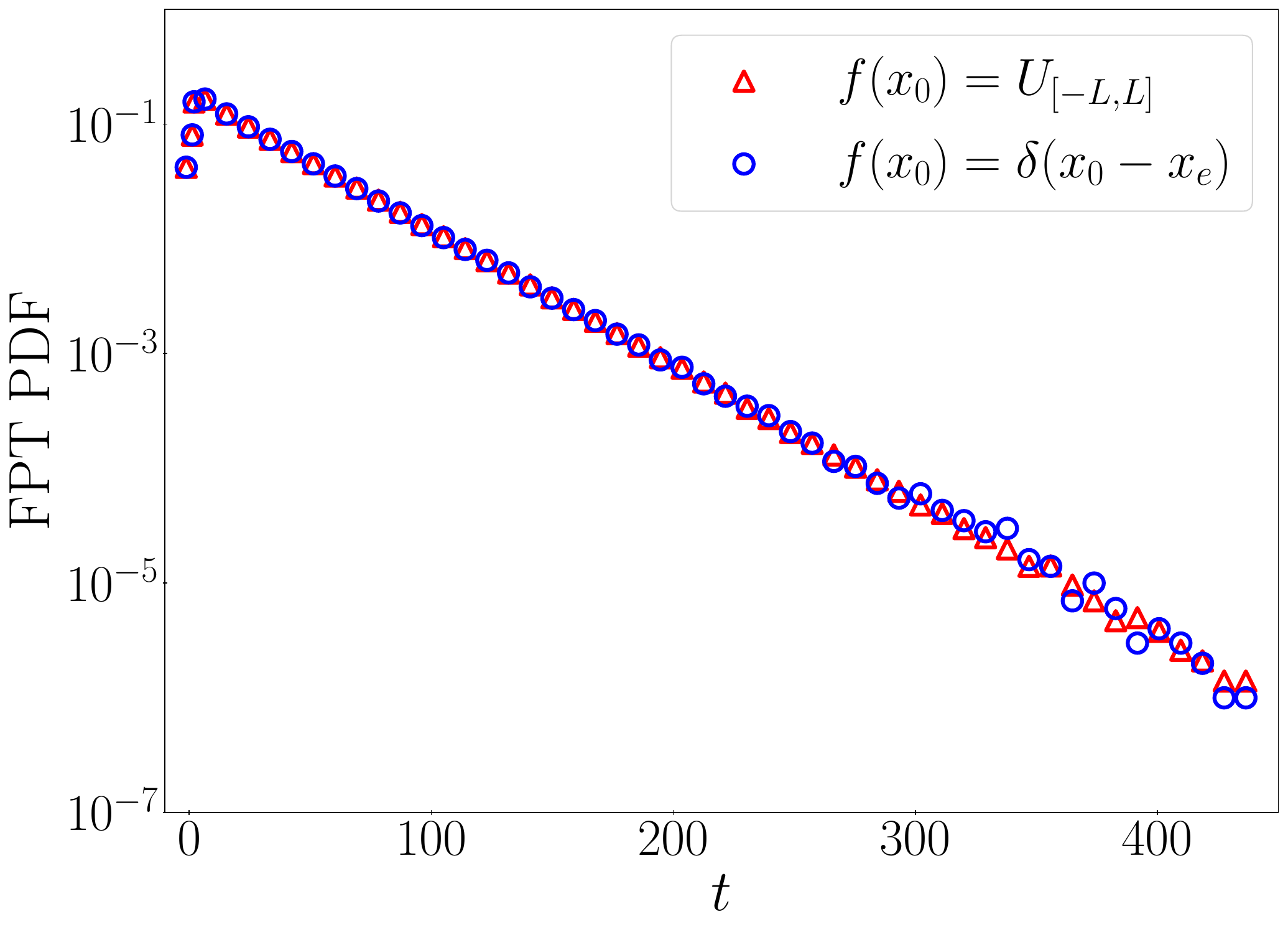}
    \caption{First-passage PDFs. We simulate the resetting process to a point $x_0$ randomly sampled from the symmetric resetting points PDF in the interval $[-L,L]$ and we compute the first-passage PDF, represented in red triangles. We then simulate the resetting process to the ERP and compute the first-passage PDF, represented in blue circles. The $y$-axis is in logarithmic scale, so we can see the exponential behavior of the PDF at long times, as expected from section IIa. The parameters used for the numerical simulations are $x_T = 5$, $D=1$, $r=0.1$, $L=2$ and $N=10^7$ trajectories.}
    \label{fig:figPDF}
\end{figure}

We address now the following question: given a specific resetting point PDF $f(x_0)$, what is the specific value of a single reset point $x_e$ such that the resulting MFPT is the same as the AMFPT obtained for the distributed case with $f(x_0)$? The value of $x_e$ can be denoted as an Equivalent Resetting Point (ERP) with respect to the PDF $f(x_0)$. As we have seen in Section \ref{universalpdf}, any choice of $f(x_0)$ which shares the same MFPT will lead to the same first-passage PDF, so we expect that the first-passage statistics of resetting to the ERP will be completely equivalent to the case of distributed resetting, independent of $f(x_0)$. In Figure \ref{fig:figPDF} we confirm this with with numerical simulations for the case where $f(x_0)$ is taken as an uniform distribution in the interval $(-L,L)$. 

Considering \eqref{ta1} and \eqref{I} with  $f(x_0)=\delta (x_0-x_e)$ the MFPT for the ERP will be of the general form
\begin{eqnarray}
  \left\langle T_{e}\right\rangle _{x_{0}}=\frac{1}{r}\left(e^{|x_{e}-x_{T}|\sqrt{\frac{r}{D}}}-1\right).
 \label{te}
\end{eqnarray}
In consequence, by the definition of $x_e$, equating \eqref{te} to \eqref{ta1} the following relation has to be satisfied
\begin{eqnarray}
    |x_{e}-x_{T}|=-\sqrt{\frac{D}{r}}\ln\left[I(r,x_{T})\right].
    \label{defxe}
\end{eqnarray}
As we have proven in the Appendix, $I(r,x_T)<1$ so that the sign of the right-hand side of Eq. \eqref{defxe} is positive. Solving Eq. 
\eqref{defxe} for $x_e$ we find two possible solutions:
\begin{eqnarray}
    x_{e}^\pm=x_T\mp \sqrt{\frac{D}{r}}\ln\left[I(r,x_{T})\right].
    \label{e1}
\end{eqnarray}
Both solutions for the ERP are at the same distance from the target, the positive (negative) solution corresponding to the ERP at the right (left) side of the target, respectively. 

Let us consider now some particular cases. The ERP corresponding to the symmetric resetting points PDF in the interval $[-L,L]$ with $x_T\notin \mathcal{D}$, is from \eqref{12} and \eqref{e1}

\begin{eqnarray}
    x_e^+&=&2x_T-\sqrt{\frac{D}{r}}\ln\left[\int_{-L}^{L}f(x_0)e^{x_0\sqrt{\frac{r}{D}}}dx_0\right]\nonumber\\
    x_e^-&=&\sqrt{\frac{D}{r}}\ln\left[\int_{-L}^{L}f(x_0)e^{x_0\sqrt{\frac{r}{D}}}dx_0\right].
    \label{xe}
\end{eqnarray} 
For an uniform resetting points PDF in $[-L,L]$ with $x_T>L$, and using \eqref{xe}, we obtain
\begin{eqnarray}
        x_e^+&=&2x_T-\sqrt{\frac{D}{r}}\ln\left[\frac{1}{L}\sqrt{\frac{D}{r}}\sinh\left(L\sqrt{\frac{r}{D}}\right)\right]\nonumber\\
    x_e^-&=&\sqrt{\frac{D}{r}}\ln\left[\frac{1}{L}\sqrt{\frac{D}{r}}\sinh\left(L\sqrt{\frac{r}{D}}\right)\right].
    \label{xeu}
\end{eqnarray}

Alternatively, for the case in Eq. \eqref{2d} where the searcher resets to $-L$ or to $L$ with probabilities $1-p$ and $p$, with $x_T>L$, using Eq. \eqref{e1} we get
\begin{eqnarray}
        x_e^+&=&2x_T-\sqrt{\frac{D}{r}}\ln\left[pe^{L\sqrt{\frac{r}{D}}}+(1-p)e^{-L\sqrt{\frac{r}{D}}}\right]\nonumber\\
    x_e^-&=&\sqrt{\frac{D}{r}}\ln\left[pe^{L\sqrt{\frac{r}{D}}}+(1-p)e^{-L\sqrt{\frac{r}{D}}}\right]
    \label{e2}
\end{eqnarray}

\begin{figure}[htbp]
    \includegraphics[width=\hsize]{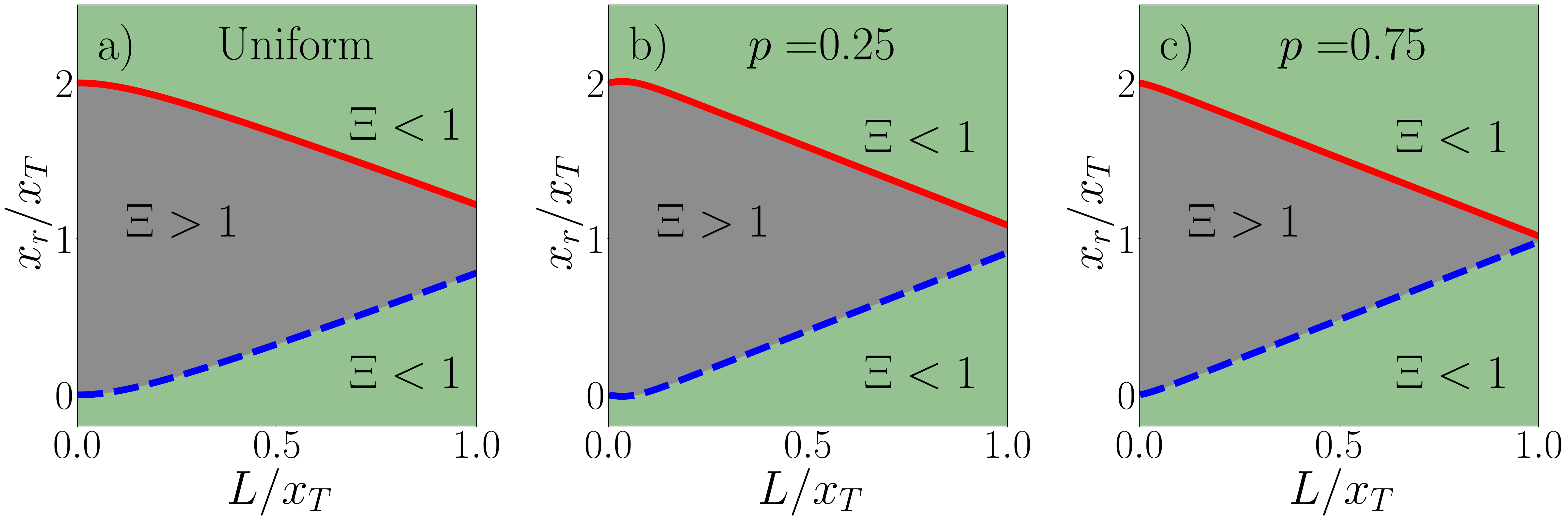}
    \caption{Diagram $x_r/x_T$ vs $L/x_T$ for the equivalent resetting point with $x_T\notin \mathcal{D}$.  We have obtained the value of $\Xi$, defined in Eq. \eqref{eq:xi}, numerically using Eq. \eqref{EM} and Eq. \eqref{Ts} in panel a) and Eq. \eqref{EM} and Eq. \eqref{t27} in panels b) and c). In the green region ($\Xi>1$), it is more efficient to reset to distributed random points. However, in the grey region, the efficient strategy is to reset to $x_r$. The red solid line represents $x_e^+$ while the blue dashed line represents $x_e^-$. Panel a) corresponds to a uniform resetting points PDF in $[-L,L]$ the red and blue lines being computed with Eq. \eqref{xeu}. Panels b) and c) the searcher resets to $-L$ or $L$ with probabilities $1-p$ and $p$, respectively. Then the red and blue lines are computed with Eq. \eqref{e2}. Panel b) is for $p=0.25$ and panel $c)$ is for $p=0.75$.}
    \label{fig:fig9}
\end{figure}
When $x_T\in \mathcal{D}$ the solutions given by Eq. \eqref{e1} can be written as
\begin{eqnarray}
    x_{e}^+&=&x_T-\Delta(x_T) \nonumber\\
    x_e^-&=&x_{T}+\Delta(x_T) 
\label{xe21}
\end{eqnarray}
where $\Delta(x_T)=\sqrt{D/r}\ln [I(r,x_T)]$.
If we consider the uniform resetting points PDF in $[-L,L]$ with $-L<x_T<L$,  the integral $I(r,x_T)$ is 
given by \eqref{I4} and 
\begin{eqnarray}
 \Delta(x_T)=\sqrt{\frac{D}{r}}\ln\left\{ \frac{1}{L}\sqrt{\frac{D}{r}}\left[1-e^{-L\sqrt{\frac{r}{D}}}\cosh\left(x_{T}\sqrt{\frac{r}{D}}\right)\right]\right\}.  
 \label{d1}
\end{eqnarray}
When the PDF of resetting points is given by Eq. \eqref{exp} then
\begin{eqnarray}
 \Delta(x_T)=\sqrt{\frac{D}{r}}\ln\left\{ \frac{\alpha}{\alpha^{2}-\frac{r}{D}}\left(\alpha e^{-x_{T}\sqrt{\frac{r}{D}}}-\sqrt{\frac{r}{D}}e^{-x_{T}\alpha}\right)\right\}. 
 \label{d2}
\end{eqnarray}

Finally, for the asymmetric case given by Eq. \eqref{2d} with $x_T<L$, we get
\begin{eqnarray}
 \Delta (x_T)=\sqrt{\frac{D}{r}}\ln\left[e^{-L\sqrt{\frac{r}{D}}}\left(pe^{x_{T}\sqrt{\frac{r}{D}}}+(1-p)e^{-x_{T}\sqrt{\frac{r}{D}}}\right)\right].   
 \label{d3}
\end{eqnarray}

We are now interested in comparing the AMFPT of resetting to distributed points $\left\langle T\right\rangle _{x_{0}}$ with the AMFPT of resetting to a fixed point, say $x_r$, $\left\langle T_r\right\rangle _{x_{0}}$. To this end, we define the dimensionless quantity
\begin{eqnarray}
    \Xi = \frac{\left\langle T\right\rangle _{x_{0}}}{\left\langle T_{r}\right\rangle _{x_{0}}}
    \label{eq:xi}
\end{eqnarray}
where
$
  \left\langle T_{r}\right\rangle _{x_{0}}=\left(e^{|x_{r}-x_{T}|\sqrt{\frac{r}{D}}}-1\right)/r.
$
Obviously, using the definition of the ERP, we have that $x_r=x_e$ leads immediately to $\Xi =1$. 

\begin{figure}[htbp]
    \includegraphics[width=0.8\hsize]{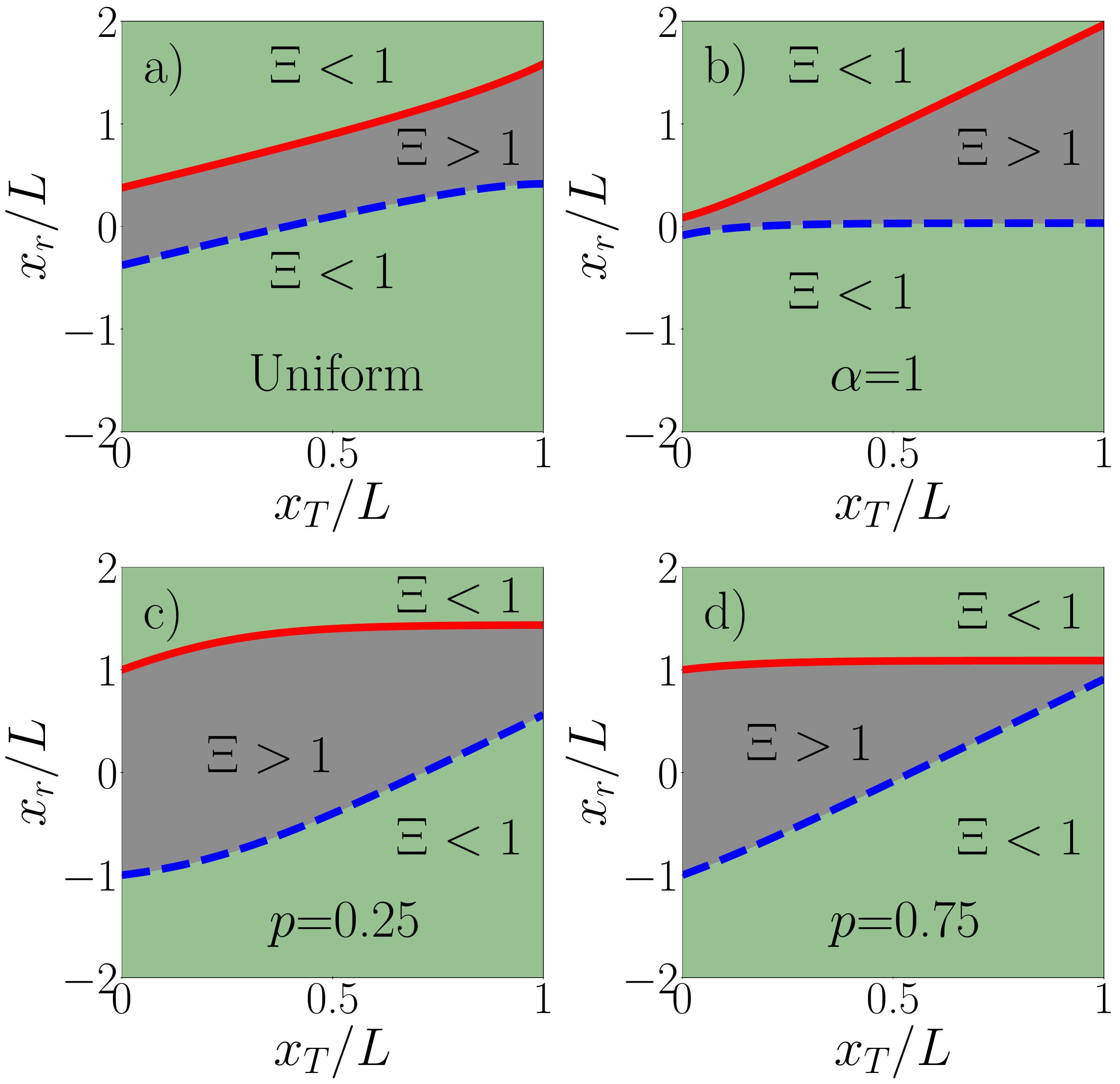}
    \caption{Diagram $x_r/L$ vs $x_T/L$ for the equivalent resetting point when $x_T \in \mathcal{D}$. We have obtained the value of $\Xi$, defined in Eq. \eqref{eq:xi}, numerically using Eq. \eqref{EM} and Eq. \eqref{ta5} in panel a), Eq. \eqref{EM} and Eq. \eqref{ta7} in panel b), and  Eq. \eqref{EM} and Eq. \eqref{T4} in panels c) and d). In the green region ($\Xi>1$), it is more efficient to reset to distributed random points. However, in the grey region, the efficient strategy is to reset to $x_r$. The red solid line represents $x_e^+$ while the blue dashed line represents $x_e^-$ computed with Eq. \eqref{xe21}. We make use of \eqref{d1} in panel a), \eqref{d2} in panel b) and \eqref{d3} in panels c) and d).} 
    \label{fig:fig10}
\end{figure}

In Figures \ref{fig:fig9} and \ref{fig:fig10}
we illustrate a comparison between both resetting protocols. In the former we plot the diagram $x_r/x_T$ versus $L/x_T$ for the equivalent resetting point when $x_T\notin \mathcal{D}$ (these curves have been plotted from Eq. \eqref{e2}). In the latter we show a diagram for $x_r/x_T$ versus $L/x_T$ s when $x_T\in \mathcal{D}$ (these curves correspond to the results found in Eq. \eqref{xe21} together with \eqref{d1}, \eqref{d2} and \eqref{d3}).

\section{Fluctuations of the MFPT}
To complete our discussion about the AMFPT, in this Section we will study the corresponding fluctuations for the case of a Brownian searcher with distributed resetting.

Averaging \eqref{exppf}, \eqref{expq} and \eqref{rela} over the resetting points, we have
\begin{eqnarray}
  \left\langle T^{n}\right\rangle _{x_{0}}=(-1)^{n+1}n\left[\frac{\partial^{n-1}\left\langle Q(s)\right\rangle _{x_{0}}}{\partial s^{n-1}}\right]_{s=0},\quad n\geq1.  
\label{Tnp}
\end{eqnarray}
Hence, for $n=2$ the second moment $\left\langle T^{2}\right\rangle _{x_{0}}$ is given by
\begin{eqnarray}
\left\langle T^{2}\right\rangle _{x_{0}}&=&\frac{2}{rI(r,x_{T})}\left\{ \partial_{r}I(r,x_{T})\right.\nonumber\\
&+&\left. \frac{1+r\partial_{r}I(r,x_{T})}{rI(r,x_{T})}\left[1-I(r,x_{T})\right]\right\} 
\end{eqnarray}
Accordingly, the variance of the first-passage time is
\begin{eqnarray}
    \sigma^{2}=\left\langle T^{2}\right\rangle _{x_{0}}-\left\langle T\right\rangle _{x_{0}}^{2}=\frac{1}{r^{2}}\left[\frac{1+2r\partial_{r}I(r,x_{T})}{I(r,x_{T})^{2}}-1\right]
\end{eqnarray}
and the coefficient of variation (CV), defined as the quotient between the standard deviation and the AMFPT, can be expressed as
\begin{eqnarray}
\textrm{CV}=\frac{\sigma}{\left\langle T\right\rangle _{x_{0}}}=\frac{\sqrt{1-I(r,x_{T})^{2}+2r\partial_{r}I(r,x_{T})}}{1-I(r,x_{T})}.
    \label{CV}
\end{eqnarray}
First of all we note that the condition for the optimal resetting rate, i.e,  $\partial \left\langle T\right\rangle _{x_{0}}/\partial r=0$ can be combined with \eqref{CV} to find $\textrm{CV}=1$, in agreement with \cite{Re16}, this is, $\textrm{CV}$ is universally equal to 1 at the optimal resetting rate. For the examples studied above we want to study the behavior of CV with the width of the resetting interval $L$ and also with the resetting rate $r$. 

\subsection{Example: Uniform resetting PDF with $x_T>L$}
\begin{figure}[htbp]
    \includegraphics[width=0.8\hsize]{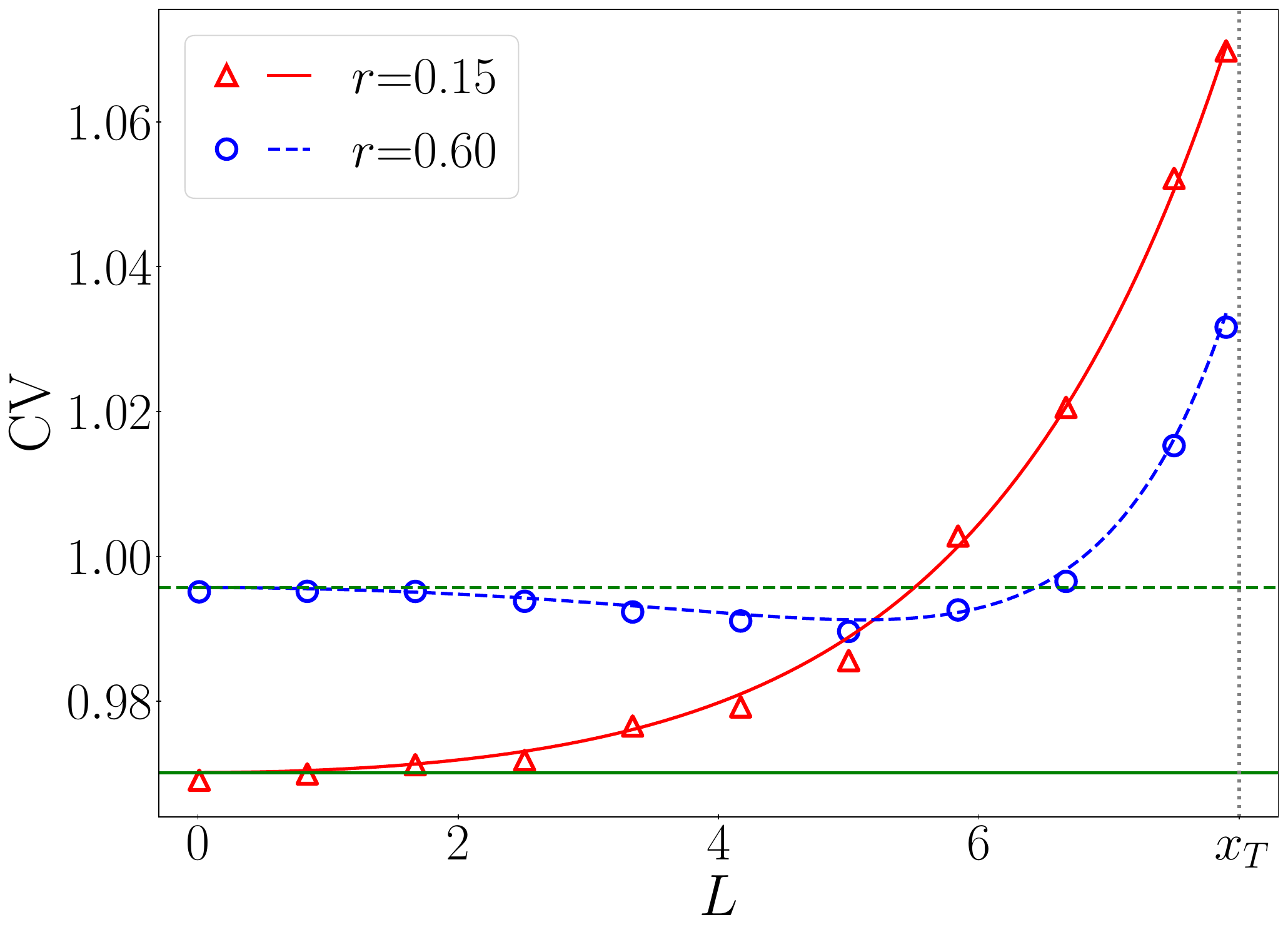}
    \caption{Coefficient of variation CV as a function of $L$ for $r=0.15$, $r<r_c$, (red solid line and triangles) and $r=0.60$, $r>r_c$ (blue dashed line and circles). The lines represent Eq. \eqref{CV} while the points are data obtained from numerical simulations. The green horizontal lines represent the EM case. The parameters used are $x_T = 8$, $D=1$ and $N=10^6$ trajectories. $r_c=0.23$.}
    \label{fig:fig11}
\end{figure}

We consider first the case when $f(x_0)$ is uniform in, $(-L,L)$, with $x_T>L$. From the definitions in \eqref{def}, together with \eqref{Iud}, we get 
\begin{eqnarray}
 I(\epsilon,z)&=&\frac{\sinh(\epsilon z)}{\epsilon ze^{z}},\quad2r\frac{\partial I(r,x_{T})}{\partial r}=z\frac{\partial I(\epsilon,z)}{\partial z}\nonumber\\
 &=&e^{-z}\left[\cosh(\epsilon z)-\frac{1+z}{\epsilon z}\sinh(\epsilon z)\right].  
 \label{ii}
\end{eqnarray}
Introducing these results into \eqref{CV} we get the expression for the CV corresponding to the uniform resetting PDF with $0<L<x_T$ (or $0<\epsilon<1$), namely $\textrm{CV}(\epsilon,z)$. For fixed $x_T$ the limit $L\to 0$, that is, $\epsilon\to 0$ we find
\begin{eqnarray}
 \textrm{CV}_{EM}=\textrm{CV}(\epsilon=0,z)=\frac{\sqrt{e^{2z}-ze^{z}-1}}{e^{z}-1},
 \label{CVEM}
\end{eqnarray}
which corresponds to the EM limit found in Eq. (3) in Ref. \cite{Re16}. On the other hand, in the opposite limit $L \to x_T$ one has $\epsilon\to 1$ so that
$$
\textrm{CV}(\epsilon=1,z)={\frac{\Phi(z)}{1+\left(2\,z-1\right){{\rm e}^{2\,z}}}},
$$
where we have defined
$$
\Phi(z) \equiv \sqrt{\left(4z^{2}+2z+2\right)e^{2z}-1+\left(4z^{2}-2z-1\right)e^{4z}}.
$$
Although it can be numerically proven that $\textrm{CV}(\epsilon=1,z)>\textrm{CV}(\epsilon=0,z)$ for any $z\in (0,\infty)$ and $\epsilon\in [0,1]$ using the results above, it does not prove in general that a minimum value for $\textrm{CV}(\epsilon,z)$ exists, for a given $\epsilon$, in this case. However, analyzing the behaviour of $\textrm{CV}(\epsilon,z)$ near $\epsilon=0$, it can be shown that $\left[\textrm{\ensuremath{\partial}CV}(\epsilon,z)/\partial\epsilon\right]_{\epsilon=0}=0$ but the second derivative $\left[\textrm{\ensuremath{\partial^{2}}CV}(\epsilon,z)/\partial\epsilon^{2}\right]_{\epsilon=0}>0$ if $z\in (0,z^*)$ and $\left[\textrm{\ensuremath{\partial^{2}}CV}(\epsilon,z)/\partial\epsilon^{2}\right]_{\epsilon=0}<0$ if $z\in (z^*,0)$, where $z^*=3.83$. Therefore, there is an optimal value $L_{opt}$ for $L$ which minimizes $\textrm{CV}(\epsilon,z)$ if and only if $r>r_c$, where $r_c=14.67D/x_T^2$. We conclude that if $r>r_c$ the fluctuations of the FPT when the searcher resets to points uniformly distributed with $L=L_{opt}$ are smaller than when resetting to a point ($L=0$). Otherwise, when $r<r_c$ the resetting process to a point has smaller fluctuations than resetting to points uniformly distributed regardless of the value of $L$. In Figure \ref{fig:fig11} we compare CV computed from (\eqref{CV}-\eqref{ii}) (solid curves) with numerical simulations (symbols), showing an excellent agreement. It is shown that for $r>r_c$ (in blue) the CV reaches a minimum value, while for $r<r_c$ (in red) such minimum does not exist. We also plot (green lines) the CV corresponding to the EM case, i.e., where the searcher resets to the origin ($L=0$), computed from \eqref{CVEM}. When $r<r_c$ we have $\textrm{CV}>\textrm{CV}_{EM}$ while if $r>r_c$ then this condition depends on $L$.

Now we want to explore how the CV depends on the reset rate $r$. To do this, we do $z\to 0$ (with the rest of values fixed) and obtain from \eqref{ii} that $\textrm{CV}\simeq z^{-1/2}\sim r^{-1/4}$, so that CV diverges as $r$ tends to 0. This is equivalent to $z\to\infty$, and from \eqref{ii} we find that CV tends to 1. In this case, although we have numerically observed that there is always a minimum for CV, we are not able to prove rigorously its existence. 

\begin{figure}[htbp]
    \includegraphics[width=0.8\hsize]{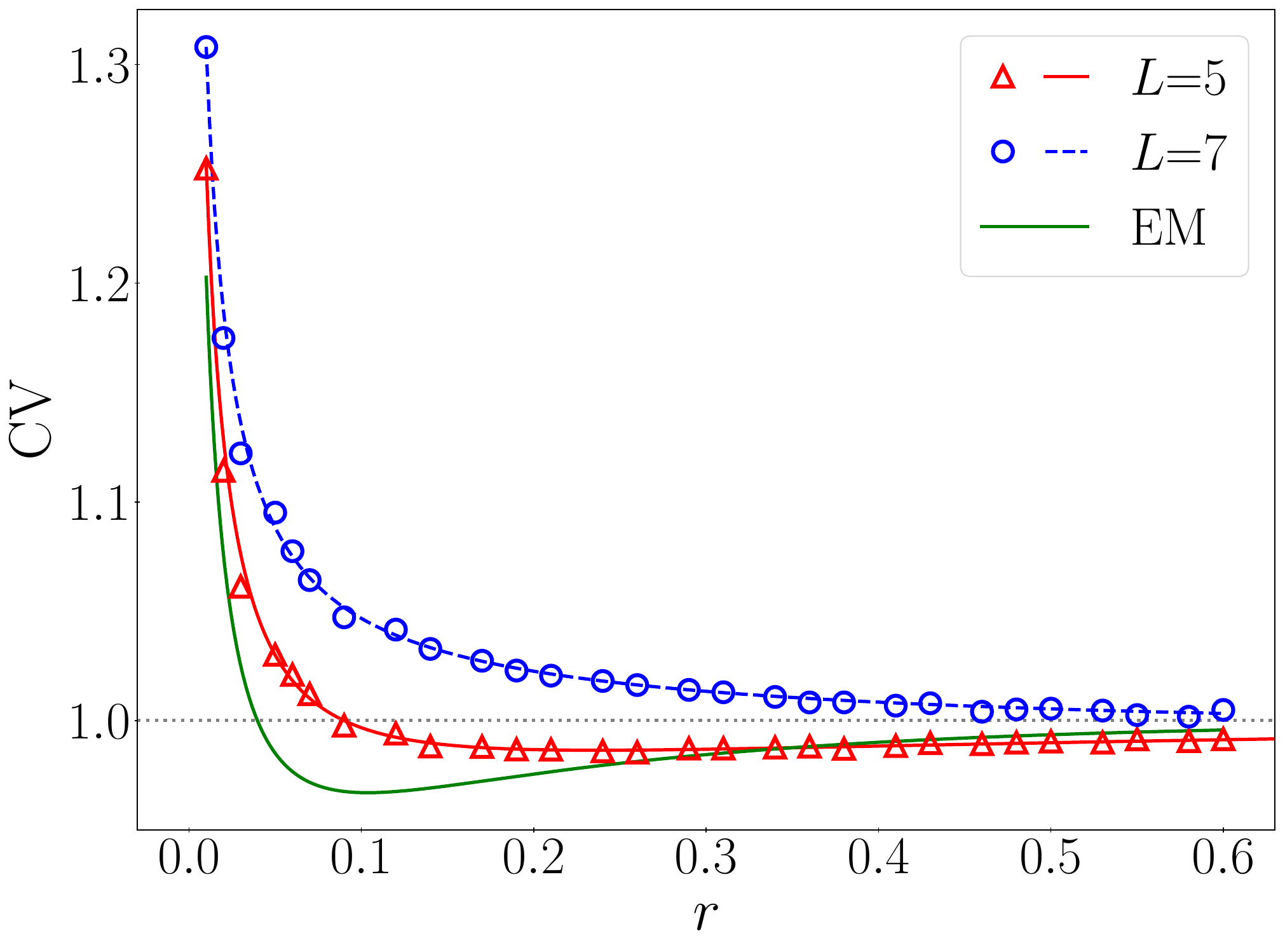}
    \caption{Coefficient of variation CV as a function of $r$ for $L=5$, (red solid line and triangles) and $L=7$ (blue dashed line and circles). The lines represent Eq. \eqref{CV} while the points are data obtained from numerical simulations. The parameters used are $x_T = 8$, $D=1$ and $N=10^6$ trajectories. The value of $r$ at which CV$=1$ is $r_{opt}=0.044$.}
    \label{fig:fig12}
\end{figure}

In Figure \ref{fig:fig12} we plot CV as a function of $r$ for two values of $L$. We show in green the case CV$_{EM}$, which allows us to see that depending on the values of $r$, CV can be higher or lower than CV$_{EM}$.   

\subsection{Example: Uniform resetting PDF with $x_T\in \mathcal{D}$}
Let us consider now the case $x_T>0$ for simplicity. Introducing the dimensionless parameters \eqref{def} into Eq.\eqref{I4} it turns into
$$
I(\epsilon,z)=\frac{1}{\epsilon z}\left[1-e^{-\epsilon z}\cosh(z)\right].
$$
Since $x_T>0$ and $-L<x_T\leq L$, then $\epsilon\in [1,\infty)$ holds. This can be used to compute CV from Eq. \eqref{CV}. To analyze the behavior of CV with $L$ we fix a value of $z$ and vary $\epsilon$. We note that CV$(\epsilon\to 1^{+},z)> 1$ and CV$(\epsilon\to \infty,z)\sim 1+1/2z\epsilon +...$. Since $\left[\partial \textrm{CV}(\epsilon,z)/\partial\epsilon\right]_{\epsilon=1}>0$, we conclude that CV reaches a maximum value for a specific value of $L$. 
\begin{figure}[htbp]
    \includegraphics[width=0.8\hsize]{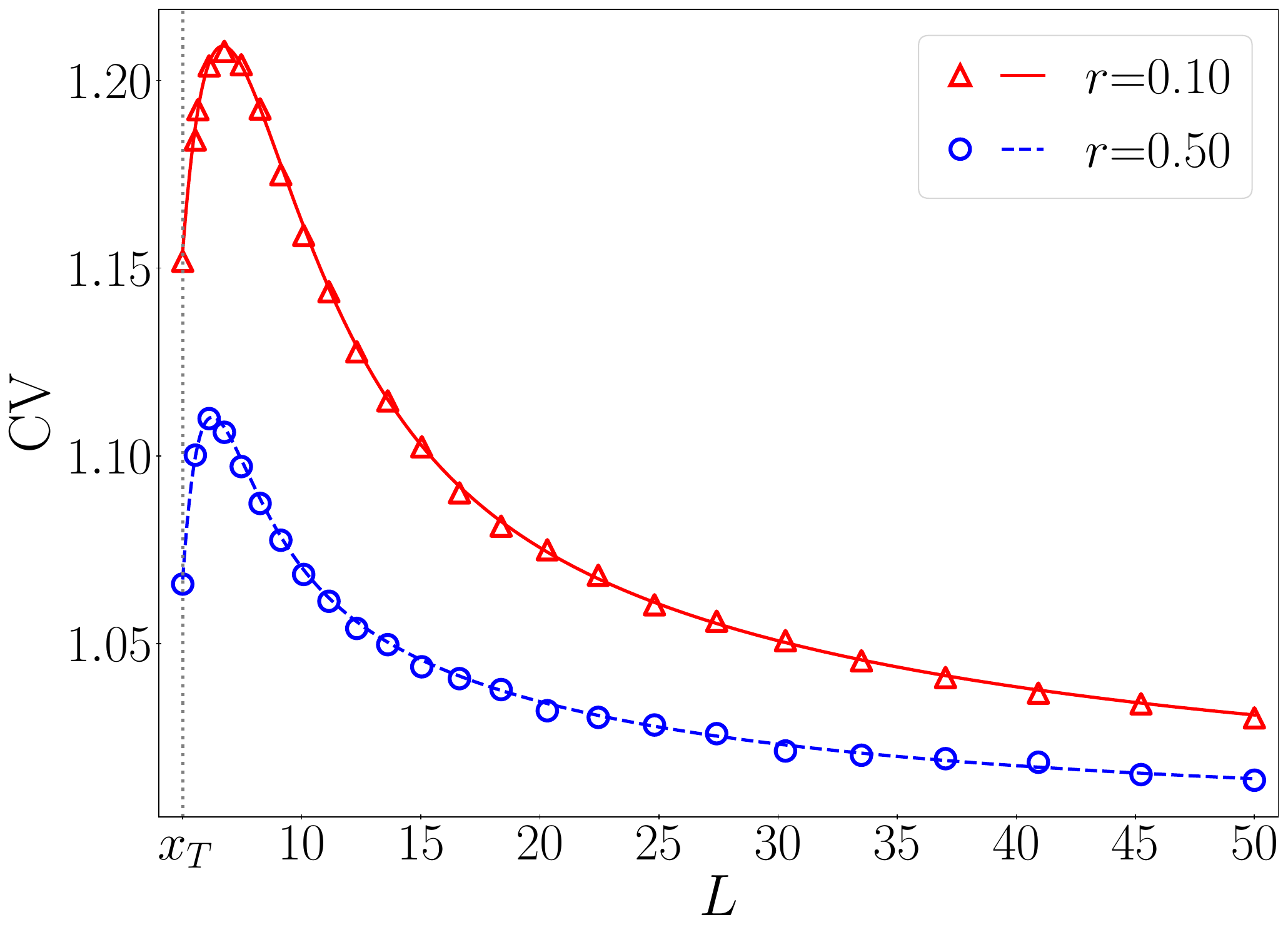}
    \caption{Coefficient of variation CV as a function of $L$ for $r=0.1$ (red solid line and triangles) and $r=0.50$ (blue dashed line and circles). The lines represent Eq. \eqref{CV} while the points are data obtained from numerical simulations. The parameters used are $x_T = 5$, $D=1$ and $N=10^6$ trajectories.}
    \label{fig:fig13}
\end{figure}

In Figure \ref{fig:fig13} we compare the CV for the uniform reset points PDF computed from \eqref{CV} with numerical simulations for two different values of $r$. It confirms the decay CV$\sim L^{-1}$ towards CV$=1$. On the other hand, if we fix $L$ and vary $r$ we see that CV$\sim r^{-1/4}$ as $r\to 0$ and CV$\sim 1$ as $r\to \infty$. Moreover, we have numerically checked that $\left[\partial \textrm{CV}/\partial r\right]<0$, so that CV decreases monotonically with $r$ towards CV $=1$. So, there is no value for $r$ such that CV$=1$, which is in agreement with the fact that there is no optimal reset rate that minimizes the AMFPT. 
\begin{figure}[htbp]
    \includegraphics[width=0.8\hsize]{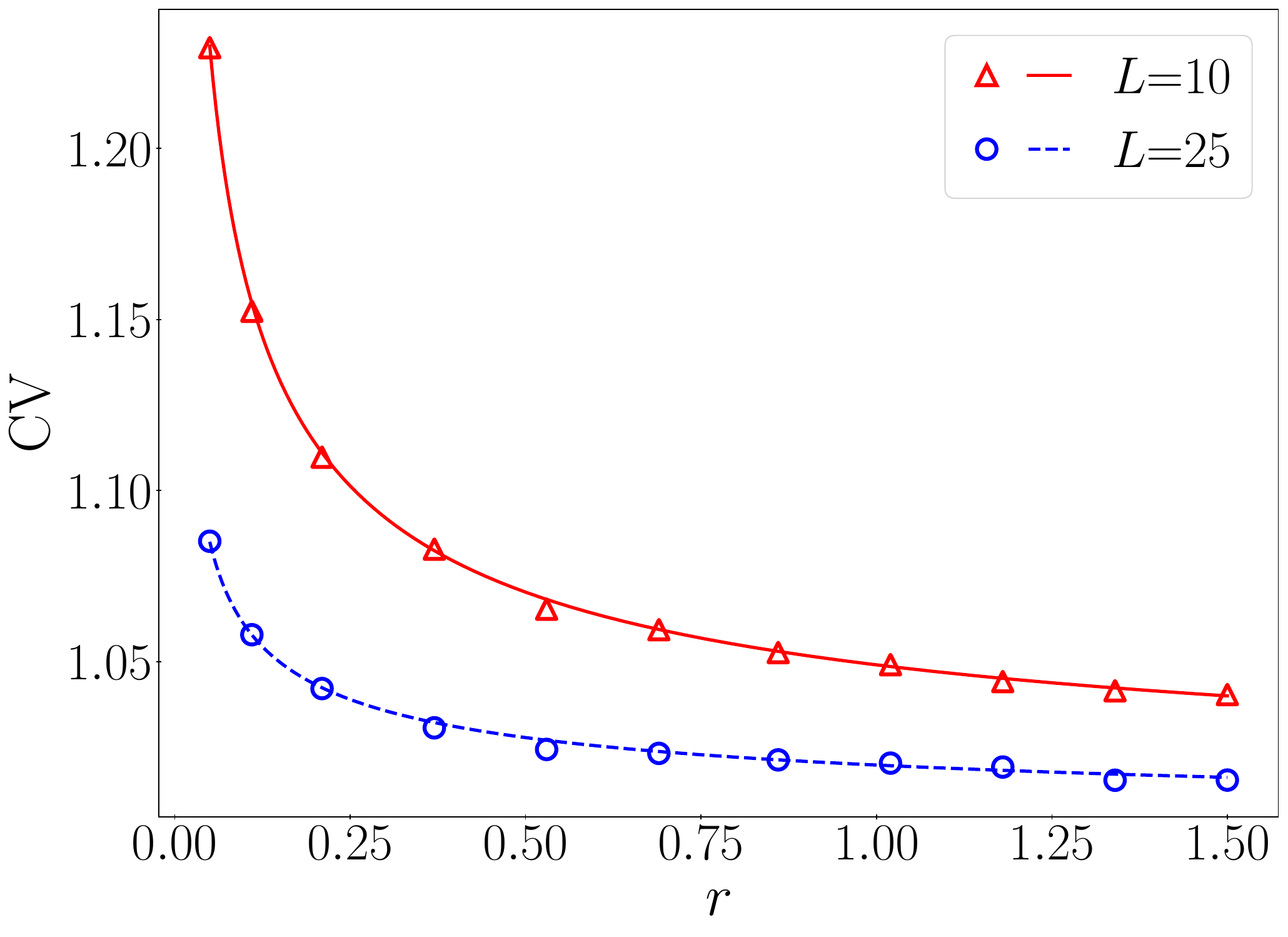}
    \caption{Coefficient of variation CV as a function of $r$ for $L=10$ (red solid line and triangles) and $L=25$ (blue dashed line and circles). The lines represent Eq. \eqref{CV} while the points are data obtained from numerical simulations. The parameters used are $x_T = 5$, $D=1$ and $N=10^6$ trajectories.}
    \label{fig:fig14}
\end{figure}
In Figure \ref{fig:fig14} we compare the CV computed from \eqref{CV} for the uniform reset points PDF with numerical simulations2, confirming the scaling of CV with $r$.

\section{Conclusions}
We have studied the effect of distributed resetting points PDF on the AMFPT of a Brownian searcher moving in one dimension. We have compared the search efficiency of resetting to the origin with the efficiency of resetting to randomly, both symmetrically or asymmetrically, distributed points around the origin. 

We have proved the existence of an optimal reset rate, which minimizes the AMFPT, when the resetting points are symmetrically distributed in an interval if the target position is outside the interval. In this case, we have shown that resetting to the origin is always less efficient than resetting uniformly on an interval. When the resetting points are symmetrically distributed but the target is located inside the domain of the resetting points PDF, then we prove that there is no optimal reset rate but there is still an optimal interval width or an optimal characteristic resetting length which minimizes the AMFPT. 

We have also considered an asymmetric resetting mechanism with different probabilities to two reset points located at the same distance to the left and right from the origin. For this case we have proved the existence of an optimal reset rate, except when the target is located on one of the resetting points. In addition, we show that the optimal reset rate diverges as a power law as the target position approaches the position of a resetting point. Additionally, we have shown that the first passage PDF averaged over the resetting points depends on the AMFPT only, revealing a somewhat universal behavior. 

Finally, we have studied the fluctuations of the MFPT as a consequence of resetting to randomly distributed points and observe that they can be higher or lower than the fluctuations of the FPT of resetting to the origin depending on the reset rate and the resetting interval width. In all cases, our theoretical predictions have been successfully compared to numerical simulations.  

\section*{Acnowledgements}
The Authors acknowledge the financial support of the Spanish government under grant PID2021-122893NB-C22

\section*{Appendix}
We consider two different situations, firstly $x_0\in \mathcal{D}$ with $x_T\notin \mathcal{D}$ but $x_T>0$ and secondly  $x_T\in \mathcal{D}$. In the former case, since $x_T>x_0$, $\forall x_{0}\in\mathcal{D}$, we have $|x_0-x_T|=|x_T-x_0|=x_T-x_0$ so that
\begin{eqnarray}
I(r,x_T)&=&e^{-x_{T}\sqrt{\frac{r}{D}}}\int_{\mathcal{D}}e^{x_{0}\sqrt{\frac{r}{D}}}f(x_{0})dx_{0}\nonumber\\
&<&e^{-x_{T}\sqrt{\frac{r}{D}}}\underset{x_{0}\in\mathcal{D}}{\max}\left(e^{x_{0}\sqrt{\frac{r}{D}}}\right)\int_{\mathcal{D}}f(x_{0})dx_{0}\nonumber\\
&=&e^{-x_{T}\sqrt{\frac{r}{D}}}\underset{x_{0}\in\mathcal{D}}{\max}\left(e^{x_{0}\sqrt{\frac{r}{D}}}\right)<1.
\label{case1}
\end{eqnarray}
In the latter case let us say $\mathcal{D}=[a,b]$ and so $a<x_T<b$. Hence,
\begin{eqnarray}
I(r,x_{T})&=&\int_{a}^{x_{T}}e^{(x_{0}-x_{T})\sqrt{\frac{r}{D}}}f(x_{0})dx_{0}\nonumber\\
&+&\int_{x_{T}}^{b}e^{-(x_{0}-x_{T})\sqrt{\frac{r}{D}}}f(x_{0})dx_{0}\nonumber\\
    &<&e^{-x_{T}\sqrt{\frac{r}{D}}}\underset{x_{0}\in(a,x_{T})}{\max}\left(e^{x_{0}\sqrt{\frac{r}{D}}}\right)\int_{a}^{x_{T}}f(x_{0})dx_{0}\nonumber\\
    &+&e^{x_{T}\sqrt{\frac{r}{D}}}\underset{x_{0}\in(x_{T},b)}{\max}\left(e^{-x_{0}\sqrt{\frac{r}{D}}}\right)\int_{x_{T}}^{b}f(x_{0})dx_{0}\nonumber\\
    &=&\int_{a}^{x_{T}}f(x_{0})dx_{0}+\int_{x_{T}}^{b}f(x_{0})dx_{0}=1.
    \label{cas2}
\end{eqnarray}
If the support of $f(x_0)$ is the real line then the above proof still holds by doing $a=-\infty$ and $b=+\infty$.

\bibliography{main}

\end{document}